%% file: main.tex
\documentclass[journal,twoside]{IEEEtran}
\IEEEoverridecommandlockouts
\usepackage[noadjust]{cite}

\usepackage{amsmath,amssymb,amsfonts}
\usepackage[ruled,noline,linesnumbered,inoutnumbered]{algorithm2e}
\usepackage{amsthm}
\usepackage{enumitem}
\usepackage{graphicx}
\usepackage{textcomp}
\usepackage{xcolor}
\usepackage[colorlinks,citecolor=blue,linkcolor=blue,urlcolor=black]{hyperref}
\usepackage{booktabs}
\usepackage{cases}
\allowdisplaybreaks[4]
\usepackage{siunitx}
\usepackage{upgreek}
\usepackage{stfloats}
\usepackage{balance}
\usepackage[smallerops,cmintegrals]{newtxmath}
\usepackage{newtxtext}
\usepackage{orcidlink}
\DeclareMathAlphabet{\mathcal}{OMS}{cmsy}{m}{n}

\usepackage{algpseudocode}

\input{symbol_def.tex}

\renewcommand{\eqref}[1]{\textcolor{blue}{(\ref{#1})}}

\begin{document}
	
\title{On Pareto-Optimal Estimation-Information Performance Limits of MIMO Integrated\\ Sensing and Communications Systems}

\author{
	\IEEEauthorblockN{Zi-Jie Wang$^{\orcidlink{0009-0009-6702-1811}}$,}
	\IEEEauthorblockN{Xudong Wang$^{\orcidlink{0000-0002-1353-1420}}$,~\IEEEmembership{Fellow,~IEEE},} and
	\IEEEauthorblockN{Giuseppe Caire$^{\orcidlink{0000-0002-7749-1333}}$,~\IEEEmembership{Fellow,~IEEE}}
	
	\thanks{Zi-Jie Wang is with the Global College, Shanghai Jiao Tong University, Shanghai 200240, China, and also with the Department of Electrical and Computer Engineering, National University of Singapore, Singapore 117583 (e-mail: zijie.wang@u.nus.edu).}
	\thanks{Xudong Wang is with the Internet of Things Thrust, The Hong Kong University of Science and Technology (Guangzhou), Guangzhou 511453, Guangdong, China (e-mail: wxudong@ieee.org).}
	\thanks{Giuseppe Caire is with the Chair of Communications and Information Theory, Technical University of Berlin, 10623 Berlin, Germany (e-mail: caire@tu-berlin.de).}
%	\thanks{Digital Object Identifier}
}

\markboth{IEEE transactions on information theory}{WANG \lowercase{et al.}: Pareto-Optimal Estimation-Information Performance Limits of MIMO ISAC}

\maketitle

\input{abstract}
\input{intro.tex}
\input{model.tex}
\input{limits.tex}
\input{SC_opt_point_result.tex}
\input{SC_opt_point_discussion.tex}
\input{numerical_result.tex}

\input{conclusion.tex}

\input{Appendix/Appendix1.tex}

\input{Appendix/Appendix2.tex}

\bibliographystyle{IEEEtran}
\bibliography{ref}
\balance
\input{bio.tex}

\end{document}

%% file: symbol_def.tex
\theoremstyle{definition}
\newtheorem{theorem}{Theorem}%[section]

\newtheorem{corollary}{Corollary}
\newtheorem{remark}{Remark}
\newtheorem{proposition}{Proposition}
\newtheorem{definition}{Definition}

\newcommand{\Zc}{\mathbf{Z}_{\rm{c}}}
\newcommand{\Zs}{\mathbf{Z}_{\rm{s}}}
\newcommand{\X}{\mathbf{X}}
\newcommand{\XX}{\overline{\X}}
\newcommand{\G}{\mathbf{G}}
\newcommand{\Y}{\mathbf{Y}}
\newcommand{\Sc}{\mathbf{S}}
\newcommand{\HH}{\mathbf{H}}
\newcommand{\R}{\mathbf{R}}
\newcommand{\F}{\mathbf{F}}
\newcommand{\N}{\mathbf{N}}

\newcommand{\Xr}{\boldsymbol{X}}
\newcommand{\XXr}{\overline{\Xr}}
\newcommand{\Yr}{\boldsymbol{Y}}
\newcommand{\Hr}{\boldsymbol{H}}

\newcommand{\Rr}{\boldsymbol{R}}
\newcommand{\RXr}{\Rr_{\Xr}}
\newcommand{\Fr}{\boldsymbol{F}}

\newcommand{\I}{{\boldsymbol{I}}}
\newcommand{\U}{\boldsymbol{U}}
\newcommand{\Ur}{\U_{\rm r}}
\newcommand{\Ut}{\U_{\rm t}}
\newcommand{\UH}{\U_{\!\Hr}}
\newcommand{\EIGr}{\boldsymbol{\mathit\Lambda}_{\rm r}}
\newcommand{\EIGt}{\boldsymbol{\mathit\Lambda}_{\rm t}}
\newcommand{\cov}{\boldsymbol{\mathit\Sigma}_{\g}}
\newcommand{\covt}{\boldsymbol{\mathit\Sigma}_{\rm{t}}}
\newcommand{\covr}{\boldsymbol{\mathit\Sigma}_{\rm{r}}}
\newcommand{\powers}{\boldsymbol{P}_{\rm s}}
\newcommand{\powerc}{\boldsymbol{P}_{\rm c}}
\newcommand{\stiefel}{\boldsymbol{\mathrm{\Psi}}}

\newcommand{\zs}{\mathbf{z}_{\rm{s}}}
\newcommand{\x}{\mathbf{x}}

\newcommand{\g}{\mathbf{g}}
\newcommand{\s}{\mathbf{s}}

\newcommand{\xr}{\boldsymbol{x}}
\newcommand{\yr}{\boldsymbol{y}}
\newcommand{\hr}{\boldsymbol{h}}

\newcommand{\uu}{\boldsymbol{u}}
\newcommand{\eigr}[1]{\lambda_{{\rm r},#1}}
\newcommand{\eigt}[1]{\lambda_{{\rm t},#1}}
\newcommand{\f}{\bar{f}}
\newcommand{\Ms}{r_{\rm s}}
\newcommand{\Ma}{r_{\alpha}}

\newcommand{\e}{{\rm{e}}}
\newcommand{\Rc}{\Rate_{\rm c}}
\newcommand{\Rs}{\Rate_{\rm s}}
\newcommand{\Ec}{\Error_{\rm c}}
\newcommand{\Es}{\Error_{\rm s}}
\newcommand{\Nc}{N_{\rm{c}}}
\newcommand{\Ns}{N_{\rm{s}}}

\newcommand{\Rate}{\mathscr{R}}
\newcommand{\Error}{\mathscr{E}}
\newcommand{\Ps}{\mathcal{P}_{\rm s}}
\newcommand{\Pc}{\mathcal{P}_{\rm c}}
\newcommand{\Po}{P_0}
\newcommand{\jj}{\jmath}

\newcommand{\st}{\rm{s.t.}}

\newcommand{\diag}{{\rm diag}}
\newcommand{\tr}{{\rm{tr}}}
\newcommand{\rank}{{\rm{rank}}}
\newcommand{\trans}{{\mathsf T}}
\newcommand{\herm}{\dagger}
\newcommand{\diff}{\mathrm{d}}

\newcommand{\CRB}{Cram\'{e}r-Rao bound}
\newcommand{\PX}{p_{\X|\HH}(\Xr|\Hr)}

\newcommand{\PXopt}{p_{\X|\HH,\alpha}^{\star}(\Xr|\Hr)}

\newcommand{\PY}{p_{\Y|\HH}(\Yr|\Hr)}

\newcommand{\PYopt}{p_{\Y|\HH,\alpha}^{\star}(\Yr|\Hr)}

%% file: abstract.tex
\begin{abstract}
Integrated sensing and communications (ISAC) holds promise for next-generation networks. Characterizing its fundamental limits is a key focus in information theory. Existing work either presumes Gaussian ISAC signals (waveforms) or approximates limits by combining sensing and communication optimum-achieving strategies, yielding loose bounds and leaving the true limits insufficiently characterized. This paper considers a general MIMO ISAC system, wherein a common ISAC signal is transmitted to convey information over the communication channel and serves simultaneously as a reference signal for sensing channel estimation. In particular, achievable data rate and channel estimation error are adopted as communication and sensing metrics, respectively. Since the transmitted waveform must be random and its ISAC-optimal distribution is unknown, the communication rate is formulated via the primitive mutual information definition, instead of the log-determinant expression under Gaussian distribution assumptions; the sensing metric is the expectation of channel estimation minimum mean square error (MMSE) with respect to the waveform distribution, representing the average estimation performance. Both metrics are functionals of the underlying waveform distribution. Characterizing the ISAC limit thus requires multi-objective optimizations over the infinite-dimensional space of probability measures, seeking distributions to achieve the Pareto-optimal estimation-information tradeoff. Employing variational analysis, the necessary condition for optimal distributions is derived, leading to high-dimensional convolutional equations. It is proven that, except at the communication-optimal point, this condition cannot be satisfied by any continuous and fully supported distributions. This result renders the ISAC limit a theoretical supremum within this distribution class, excludes Gaussian from being ISAC-optimal, and makes characterizing the optimal distribution analytically intractable. A continuous-version Blahut-Arimoto-type numerical algorithm is thus proposed to approximate the optimal distributions, and to determine corresponding MMSE-Rate performance, with provably limit-converging properties. Furthermore, closed-form characterizations of sensing- and communication-optimal points are presented: the optimal waveforms are determined by power allocation according to channel statistics/realization and waveform randomness selection. Loose performance bounds are further established for fast evaluation of the MMSE-Rate limits. Finally, numerical examples validate the theoretical results. Notably, the structural insights for ISAC limits-achieving strategies highlight the necessity of designing singular or non-standard waveforms for optimal ISAC performance.
\end{abstract}


%% file: intro.tex
%\clearpage
\section{Introduction}
\IEEEPARstart{A}{s next-generation} wireless networks are anticipated to accommodate both high-quality sensing and communication applications, sensing and communication witness a significant shift from isolation to integration, referred to as integrated sensing and communications (ISAC). ISAC allows wireless devices to conduct data transmission and environmental sensing simultaneously, utilizing the same signals and hardware platform \cite{FanLiu6G,LiuAn,signal_processing_technique_ISAC,globecom_2023,Wang2025DeviceFreeISAC}. By leveraging the dual functionality of ISAC, future wireless networks significantly enhance their communication reliability, efficiency, and intelligence while addressing the diverse and demanding requirements of various sensing applications.

\subsection{Related Works and Research Gap}

Research on wireless sensing and communication has been conducted in parallel for decades \cite{LiuAn,signal_processing_technique_ISAC}. As landing conditions for ISAC become increasingly mature, there is a pressing need to unfold unified studies and explore the fundamental theories for ISAC. Among them, identifying the performance limits is essential for understanding the influence of channel parameters and for guiding the design of optimal (i.e., limits-achieving) signaling strategies \cite{signal_processing_technique_ISAC,LiuAn,FanLiu6G}. There have been pioneering theoretical works reported for performance limits in ISAC systems. In \cite{source_paper,globecom_2023,TVT_BA}, the sensing and communication signals are modeled independent in a multiple-access framework, sharing the same ISAC channel. Inner performance bounds are derived by allocating time, frequency, and/or power resources between them. The works \cite{ISAC_State,YaoLiu_TIT} characterize the capacity-distortion tradeoff of single-input-single-output (SISO) ISAC systems from a coding-centric perspective. These works consider abstract channels (e.g., discrete memory-less channels fully described by state transition probabilities \cite{ISAC_State,YaoLiu_TIT}), wherein sensing is modeled as a generalized feedback process for state estimation over a finite reconstruction alphabet, providing valuable insights for interpreting ISAC from rate-distortion perspectives.

In contrast, a distinct line of research \cite{CRB_hua,CRB_BF} considers explicit physical channel models, e.g., multiple-input-multiple-output (MIMO) ISAC systems. Specifically, one common MIMO ISAC waveform is transmitted for radar estimation and information transmission. They investigate the performance limits of such systems from a joint estimation-information perspective, deriving the bounds on the Cram\'{e}r-Rao-Shannon achievable region under different setups. Despite their huge success, the analysis, however, neglects the discussion of the randomness of the ISAC waveform. Consequently,
	\begin{itemize}
		\item[i)] the rate is evaluated simply using the log-determinant formula, which presumes Gaussian inputs;
		\item[ii)] the input sample correlation matrix, which is inherently random, is identified with its statistical counterpart --- an approximation only valid asymptotically for infinite channel coherence time.
	\end{itemize}
Under such simplifications, both the \CRB\ (CRB) and rate depend solely on the input correlation matrix. By optimizing the weighted sum of CRB and rate with respect to the correlation matrix through standard convex optimizations, these works \cite{CRB_hua,CRB_BF} obtain certain performance bounds under different setups. 

\begin{figure}[t]
	\centering
	\includegraphics[scale=1]{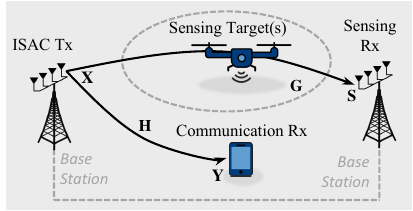}
	\caption{A notional MIMO ISAC system.
		Plotted is bi-static sensing for demonstration purposes. The sensing Rx can be colocated with the ISAC Tx for monostatic sensing.}
	\label{fig:ISAC_Setup}
\end{figure}
	
Unlike conventional communication-only systems wherein Gaussian waveforms are rate-optimal (under additive Gaussian noise), ISAC signals serve sensing purposes in addition to solely communicating. Therefore, the optimal distributions of ISAC waveforms are unknown and Gaussian-optimality may potentially break down for ISAC. As the distribution-level degrees of freedom have not been explored, the bounds derived in \cite{CRB_hua,CRB_BF} are intrinsically loose. In fact, the random nature of ISAC signals reveals a paradigm shift from second-order statistics design to distribution level characterization of the input waveform, which requires the ISAC limits to be represented by \emph{functional optimizations}. In the foundational framework \cite{FanLiu}, signal randomness is incurred for fundamental tradeoffs and tighter bounds in MIMO ISAC systems. Particularly, the ISAC signal in \cite{FanLiu} is regarded as a random reference signal for sensing. Their work highlights that the sensing and communication tradeoff primarily stems from the degree of randomness of the transmitted ISAC signals and the subspace overlap between sensing and communication channels, paving the way for sensing- and communication-optimal waveform design in ISAC systems. However, due to the lack of closed-form expressions for rate and the high complexity of evaluating sensing metrics under random waveform (with arbitrary distributions), it remains significantly challenging to solve the functional optimizations for ISAC limits. While \cite{FanLiu} smartly approximates ISAC limits by combining sensing- and communication-optimal strategies, this approach provides loose bounds, and identifying the true ISAC limit itself remains an ongoing challenge. 

\subsection{System Setup and Principal Theoretical Contributions}

To investigate the performance limits of MIMO ISAC systems, the following model is considered in this paper \cite{FanLiu,CRB_BF,CRB_hua}
\begin{align*}
	\Y=\HH \X+\Zc;\\
	\Sc=\G \X+\Zs,
\end{align*}
where the dual-functional ISAC signal/waveform $\X$ is emitted from the ISAC transmitter (Tx), as shown in Fig. \ref{fig:ISAC_Setup}. Specifically, $\X$ is used to convey information over the communication channel $\HH$ from the received communication signal $\Y$, and to estimate the sensing channel $\G$ from the sensing signal $\Sc$ received by the sensing receiver (Rx). $\Zs$ and $\Zc$ represent the respective additive Gaussian noise into channels. Additionally, $\HH$ and $\G$ are independent (and hence,  uncorrelated); $\Zs$ and $\Zc$ are independent. Additionally, in ISAC systems, each symbol is random. 

The sensing performance is evaluated using the defined \textit{average MMSE} ($\Error$), which measures the average channel estimation performance for the random input $\X$. More specifically, it is assumed that any realization of $\X$, denoted as $\Xr$, is perfectly known at the sensing Rx.\footnote{This assumption is widely adopted in state-of-the-art \cite{FanLiu,CRB_BF,CRB_hua,Wang2025DeviceFreeISAC} to align with radar sensing models \cite{sensing_channel_assumption1,MIMO-Radar-waveform,Sensing_channel_model}, and can be practically justified with colocated (e.g., monostatic) or collaborative bistatic radar settings via the backhaul links, where $\Xr$ serves as a reference signal for sensing.} The overall evaluation of sensing performance (i.e., the average MMSE) involves taking the expectation of MMSE conditioned on $\Xr$ over each $\X=\Xr$ according to the density $p_{\X}(\Xr)$,\footnote{This trick can be justified in definition of the Miller-Chang type CRB \cite{Miller1978MCRB}, the law of iterative expectation, coherent symbol detection \cite[Section 3.1]{tse2005fundamentals}, etc., where a random nuisance parameter is involved in evaluating the overall performance.} which is referred to as ``input/waveform distribution'' hereafter. Additionally, the communication performance is assessed with the ergodic coherent rate ($\Rate$), with the assumption of perfect knowledge of channel state information at the ISAC Tx (CSIT) \cite[Section 10.3]{tse2005fundamentals}.

\begin{figure}[t]
	\centering
	\includegraphics[scale=1]{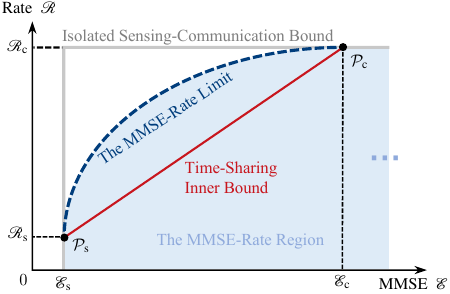}
	\caption{The MMSE-Rate region for MIMO ISAC systems: an illustrative plot.}
	\label{fig:Illustration}
\end{figure}

Based on the system configuration mentioned above, both the average MMSE and the rate are functionals of the waveform distribution $p_{\X}(\Xr)$. As the core focus of this paper, the Pareto-optimal MMSE-Rate limit, as elaborated in Fig.~\ref{fig:Illustration}, is represented by a family of optimizations over the space of distributions. More specifically, each point on this limit corresponds to a specific distribution of $\X$ that achieves a unique balance between the estimation MMSE and the rate. The key objective of this paper is to characterize the Pareto boundary of the MMSE-Rate limit, and to investigate how the distributions of $\X$ relate to achieving it. This paper's principal theoretical results and contributions are summarized as follows.
\begin{itemize}
	\item The MMSE-Rate limit is investigated in this paper.
	The Euler-Lagrange (i.e., necessary) conditions for the optimal distributions (e.g., $p_{\X}(\Xr)$ and $p_{\Y}(\Yr)$) to reach the limit are derived employing a variational calculus approach, which can be expressed as complex high-dimensional convolutional equations. It is further shown that, except at the communication-optimal point $\Pc$, the MMSE-Rate limit is unachievable over the space of standard input distributions (e.g., Lebesgue-continuous and fully-supported). This provides structural implications of the limits-achieving ISAC waveform: at any point on the Pareto limit other than $\Pc$, the optimal distribution (if it exists) must be discontinuous and/or supported on a proper subset (e.g., manifold) of the full space; specifically, at the sensing-optimal point $\Ps$, the input sample correlation matrix must be unique and deterministic.

	\item To address the general unachievability of the ISAC MMSE-rate limit, this paper develops a numerical algorithm from the variational condition. It extends the classical Blahut-Arimoto (B-A) framework \cite{Blahut,Arimoto} to the considered setup, enabling i) to determine a well-behaved input distribution converging to the MMSE-Rate limit and ii) to compute MMSE-Rate performance along the limit curve to any desired level of precision. Additionally, its convergence to the MMSE-Rate limit is proven.

	\item Closed-form characterizations of $\Ps$ and $\Pc$ are presented. The optimum-achieving waveforms are derived, as completely determined by the respective channel (statistics/realizations) and
	waveform characteristics employed; the MMSE and rate at $\Ps$ and $\Pc$ are provided in closed-form analytical expressions. This paper highlights the water-filling tradeoff (WFT) and the waveform uncertainty tradeoff (WUT) in ISAC, respectively relying on the power allocation and the selection of waveform randomness for the ISAC signal. Leveraging WFT and WUT, this paper revisits \cite{FanLiu} and establishes several loose bounds for rough evaluations of the MMSE-Rate limit.
\end{itemize}

Additionally, numerical examples are provided to illustrate and validate the derived analytical theories. To the best of the authors’ knowledge, this is the first work to investigate the fundamental estimation-information performance limits of MIMO ISAC systems in the distribution level, by considering general and random ISAC waveforms. It reveals how each operating point on the ISAC Pareto boundary, which correspond to different task-oriented preferences between sensing and communication, is fundamentally governed by the statistical properties of ISAC waveforms. In essence, it characterizes the channel capacity and capacity-achieving waveforms under explicit sensing performance constraints, beyond conventional power-limited rate optimization frameworks \cite{telatar1999capacity}. The core findings demonstrate that conventional waveforms may fail to attain the optimal ISAC performance, and singular or non-standard waveform designs are essential for optimal ISAC operation. This not only delivers practical guidelines for high-performance ISAC waveform engineering, but hopefully also motivates further in-depth analytical investigations into the rigorous characterization of ISAC fundamental limits and corresponding achieving strategies.

%The approach readily adapts to ISAC systems with various metrics, such as the estimation rate \cite{MIMO-Radar-waveform,TVT_BA} and the Miller-Chang type CRB \cite{Wang2025DeviceFreeISAC,Miller1978MCRB,FanLiu}, as well as to a broad class of channel models.

\subsection{Notation}
Notational convention throughout this paper is as follows, unless otherwise specified. Normal font indicates a random variable, as $\mathsf{x}$, $\x$, and $\X$ respectively represent a random scalar, column vector, and matrix; while their deterministic observations or realizations are italic, as $x$, $\xr$, and $\Xr$. $(\cdot)^*$, $(\cdot)^\trans$, and $(\cdot)^\herm$ are respectively the matrix conjugate, transpose, and Hermitian transpose operators.
$\rank(\cdot)$, $\tr(\cdot)$, and $\det(\cdot)$ respectively denote the rank, the trace, and the determinant of their matrix argument. $(x)^+:=\max\{0,x\}$. $\mathbb{E}(\cdot)$ is the expectation operator. ${\I}_N$ denotes identity matrix with rank $N$. $\mathbf{0}_{M\times N}$ denote the $M\times N$ matrix whose entries are all 0's. The dimensions and subscripts are neglected wherever there is no confusion. $\rm vec(\cdot)$ denotes a column-wise stacked vector of its argument, $\rm{diag}(\cdot)$ denotes a square matrix whose main diagonals are placed with the arguments. $\otimes$ denotes the Kronecker product operator. $\boldsymbol{A}\succeq\boldsymbol{B}$ implies $\boldsymbol{A}-\boldsymbol{B}$ is positive semi-definite. In this paper, conditioning is used in both information-theoretic and signal processing contexts. From information-theoretic perspectives, it refers to statistical dependence on a random variable and is used to define quantities such as conditional entropy or conditional distributions; from signal processing perspectives, it refers to the performance under a specific value of the conditioned variable.

\subsection{Organization}
The ISAC system model and performance metrics are introduced and defined in Section \ref{sec:model}. The main theoretical results for characterizing the MMSE-Rate region are presented in Section \ref{sec:results}. They include the characterization and identification of the MMSE-Rate limit in Sections \ref{sec:results_limit_identify} and \ref{sec:results_limit_compute}, respectively, and the investigation of the sensing and communication optimal points in Section \ref{sec:results_SAC}. While corresponding derivations and proofs are in Appendixes. Numerical results are presented in Section \ref{sec:numerical}. Finally, this paper is concluded in Section \ref{sec:conclusion}.
%\end{comment}

%% file: model.tex
\section{System Model}
\label{sec:model}

\subsection{ISAC Signal and System}

As shown in Fig. \ref{fig:ISAC_Setup}, the ISAC Tx with $N$ antennas transmits a dual-functional signal $\X\in\mathbb{C}^{N\times T}$ for data communication and environmental sensing purposes, where $T\geq N$ is the duration of communication frame as well as the sensing snapshot number (e.g., coherent processing interval).\footnote{In \cite{MIMO_capacity_no_CSI}, it is concluded that further increasing the transmit antenna number $N$ beyond $T$ does not further increase the capacity. $T\geq N$ is also assumed in \cite{MIMO-Radar-waveform,sensing_channel_assumption1} to guarantee the dimensionality requirement for channel estimation, i.e., it requires at least $N$ samples to estimate a matrix with $N$ columns. Therefore, for the sake of both sensing and communication, $T\geq N$ is required.} The communication Rx has $\Nc$ receiving antennas; the sensing Rx has $\Ns$ receiving antennas. The general ISAC system can be described as
\begin{subequations} \label{eqn:MMSE_rate_system}
	\begin{align}
		\Y&=\HH \X+\Zc; \label{eqn:com_model_coherent}\\
		\Sc&=\G \X+\Zs, \label{eqn:sensing_model_coherent}
	\end{align}
\end{subequations}
where $\Y\in\mathbb{C}^{\Nc\times T}$ is the received communication signal at the communication Rx,  $\HH\in \mathbb{C}^{\Nc\times N}$ is the wireless communication channel, $\Sc\in\mathbb{C}^{\Ns\times T}$ is the received sensing signal at sensing Rx,
$\G\in \mathbb{C}^{\Ns\times N}$ is the wireless sensing channel (target response/scattering matrix), $\Zc\in\mathbb{C}^{\Nc\times T}$ and $\Zs\in\mathbb{C}^{\Ns\times T}$ are the respective noises to ISAC communication and sensing sub-systems, respectively.

\subsection{Assumptions}
For further derivation, following key assumptions are made.
\begin{enumerate}[leftmargin=2.4\parindent]
	\item[(A1)] [Sensing Channel] The sensing channel is\footnote{In radar-centric works \cite{sensing_channel_assumption1,MIMO-Radar-waveform,Sensing_channel_model,Widely-Separated-Antennas}, only Tx-side channel correlation is considered. In contrast, this paper accounts for both Tx- and Rx-side correlations of the sensing channel. This captures more realistic channel statistics, and enables a wider range of sensing tasks, such as angle-of-arrival estimation.}
	\begin{equation}
		\label{eqn:sensing_channel}
		\G = \covr^{1/2}\N_{\G}\covt^{1/2},
	\end{equation}
		where $\covt\in\mathbb{C}^{N\times N}$ and $\covr\in\mathbb{C}^{\Ns\times \Ns}$ are respectively the Tx- and Rx-side correlations for the sensing channel, and each entry of $\N_{\G}\in\mathbb{C}^{\Ns\times N}$ follows i.i.d. zero-mean circular symmetric complex Gaussian distribution with variance $\sigma_{\mathsf g}^2$, i.e., ${\text{vec}}(\N_{\G})\sim\mathcal{CN}(\mathbf{0},\sigma_{\mathsf g}^2 \I_{N\Ns})$.
		The eigenvalue decompositions of $\covr$ and $\covt$ are
			\begin{equation}
				\label{eqn:eigen_decomosition_G}
				\begin{aligned}
					\covr = \Ur \EIGr \Ur^\herm;\\
					\covt = \Ut \EIGt \Ut^\herm,
				\end{aligned}
			\end{equation}
		where $\Ur$ and $\Ut$ are unitary matrices with $\covr$'s and $\covt$'s eigenvectors as their columns, respectively,  $\EIGr={\rm diag}(\eigr{1},\cdots,\eigr{\Ns})$ and $\EIGt={\rm diag}(\eigt{1},\cdots,\eigt{N})$ store all non-negative eigenvalues of $\covr$ and $\covt$ on their diagonals in a non-increasing manner, respectively.

	\item[(A2)] [Communication Channel] Each entry of $\HH$ follows i.i.d. zero-mean circular symmetric complex Gaussian distribution with variance $\sigma_{\mathsf h}^2$ \cite{Wireless_communication_Goldsmith}, i.e.,
	${\rm{vec}}(\HH)\sim\mathcal{CN}(\mathbf{0},\sigma_{\mathsf h}^2\I_{N\Nc})$.
	
	\item[(A3)] [Channel Fading Model] Both $\G$ and $\HH$ are block fading channels, which stay constant for a finite duration $T$, and change to a new realization every $T$ symbols in an i.i.d. manner \cite{MIMO_capacity_no_CSI}. $T$ is the \textit{coherence time} \cite{Grassmann} of both sensing and communication channels. Additionally, $\HH$ and $\G$ are independent \cite{FanLiu,CRB_hua,CRB_BF}.

	\item[(A4)] [Power Constraint] The transmitted signal $\X$ has the short-term power constraint
	\begin{equation}
		\mathbb{E}\Big\{\tr(\R_{\X})\big|\HH,\G \Big\}
 		%{{\tr}(\tilde{\Rr}_{\X})
 		\leq N\Po \quad {\rm w.p.\ }1,
		\label{eqn:power_constraint}
	\end{equation}
	where $\R_{\X}:=T^{-1}\X\X^\herm$ is the sample correlation matrix of $\X$.
	The statistical correlation matrix of $\X$ is $\tilde{\Rr}_{\X}:=\mathbb{E}(\R_{\X})$. Note that \eqref{eqn:power_constraint} should hold for each realization of $\HH$ and $\G$.

	\item[(A5)] [Noise Model] Each element from $\Zc$ and $\Zs$ is an i.i.d. zero-mean circularly symmetric complex Gaussian random variable with variance $\sigma_{\rm c}^2$ and $\sigma_{\rm s}^2$, i.e., ${\rm vec}(\Zc)\sim\mathcal{CN}(\mathbf{0},\sigma_{\rm c}^2\I_{T\Nc})$ and ${\rm vec}(\Zs)\sim\mathcal{CN}(\mathbf{0},\sigma_{\rm s}^2\I_{T\Ns})$, respectively. $\Zs$ and $\Zc$ are independent.

	\item[(A6)] The transmitted ISAC waveform $\X$ is known at the sensing Rx as a reference signal \cite{FanLiu}; the communication channel $\HH$ is known at both the ISAC Tx and communication Rx \cite{FanLiu,CRB_hua,CRB_BF}; the sensing channel statistics $\covt$ and $\covr$ are \emph{a priori} known at the ISAC Tx.
\end{enumerate}

\begin{remark}
	In (A1)-(A3), the flat block-fading MIMO channel is in focus, while the extension to frequency-selective MIMO channels is straightforward based on results from this paper.
\end{remark}

\subsection{ISAC Performance Metrics}
\subsubsection{Sensing}

The objective of sensing for an ISAC system is to estimate the sensing channel $\G$, which depends on parameters of interest (e.g., distance, Doppler, etc.). These parameters can be extracted from $\G$'s estimates employing well-designed signal processing algorithms \cite{LiuF-CRB-TSP}. Equivalently, the sensing model \eqref{eqn:sensing_model_coherent} can be rewritten as
\begin{equation}
	\s=\XX\g+\zs, 
	\label{eqn:sensing_model}
\end{equation}
where $\s=\rm{vec}({\Sc^\herm})$, $\XX=\I_{\Ns}\otimes\X^\herm$, $\g=\rm{vec}(\G^\herm)$ and $\zs=\rm{vec}(\Zs^\herm)$. Such a vectorizing operation transforms the original matrix-based estimation problem into an equivalent vector-based formulation, allowing classical estimation results \cite[Chapter 11]{kay1993fundamentals} to be readily applied. From \eqref{eqn:sensing_channel},
$$\begin{aligned}
	\g &={\rm vec}(\G^\herm) ={\rm vec}\big[(\covr^{1/2}\N_{\G}\covt^{1/2})^\herm\big] \\
	&=\big[(\covr^{1/2})^\trans\otimes\covt^{1/2}\big]{\rm vec}(\N_{\G}^\herm).
\end{aligned}$$
Hence, the correlation matrix of $\g$ can be further derived as
$$\begin{aligned}
	&\cov := \mathbb{E}(\g\g^\herm) \\
	&= \Big[(\covr^{1/2})^\trans\!\!\otimes\!\covt^{1/2}\Big] \mathbb{E}\Big[{\rm vec}(\N_{\G}^\herm){\rm vec}^\herm(\N_{\G}^\herm)\Big] \Big[(\covr^{1/2})^\trans\!\!\otimes\!\covt^{1/2}\Big]^\herm \\
	&= \sigma_{\mathsf g}^2 \covr^\trans\otimes\covt.
\end{aligned}$$

The sensing performance is the channel estimation MMSE. Conditioned on a particular realization $\XX=\XXr$, the joint distribution between $\g$ and $\s$ can be readily derived as 
\begin{equation}
	\left.\begin{bmatrix}
	\g \\
	\s
	\end{bmatrix}\right| \Big(\XX=\XXr\Big) \sim \mathcal{CN}\!\left(\mathbf{0},\begin{bmatrix}
	\cov&\cov\XXr^\herm\\
	\XXr\cov&\XXr\cov\XXr^\herm+\sigma_{\rm s}^2\I_{T\Ns}
	\end{bmatrix}\right).
	\label{eqn:joint_distribution}
\end{equation}

Note \eqref{eqn:sensing_model} is a linear model between observation $\s$ and parameter-to-be-estimated $\g$. The Bayesian MMSE estimator of $\g$, denoted as $\hat{\g}$, is the conditional mean of $\g$ given the observed data $\s$ \cite{kay1993fundamentals}. Based on \eqref{eqn:joint_distribution} and \cite[(10.32)]{kay1993fundamentals}, the MMSE estimator of $\g$, as a function of $\s$ and $\XXr$, is
\begin{equation}
    \hat{\g}\Big(\s,\XXr\Big)=\mathbb{E}\Big[\g\Big|\Big(\s,\XX=\XXr\Big)\Big]= \Big(\sigma_{\rm s}^2\cov^{-1}+\XXr^\herm\XXr\Big)^{-1}\XXr^\herm\s,
	\label{eqn:estimator}
\end{equation}
the correspondingly MMSE is \cite[(10.33)]{kay1993fundamentals}
\begin{equation*}
	\begin{aligned}
		\epsilon(\XXr) &= \mathbb{E}\left[\ \Big\|\hat{\g}\Big(\s,\XXr\Big)-\g\Big\|^2\ \bigg|\ \XX=\XXr\right]\\
		&=\tr\bigg[\Big(\cov^{-1}+\sigma_{\rm s}^{-2}\XXr^\herm\XXr\Big)^{-1}\bigg].
	\end{aligned}
\end{equation*}

The above MMSE depends on a specific realization of the ISAC waveform. Since the ISAC waveform is random (and known at the sensing Rx), its overall contribution to the sensing performance should be accounted for in an ensemble sense. To this end, the waveform $\X$ (or $\XX$) is treated as a nuisance parameter \cite[Chapter 10.7]{kay1993fundamentals}, and the sensing performance is characterized by the average MMSE:
\begin{equation} \label{eqn:sensing_metric}
	\Error := \mathbb{E}\left\{ \epsilon(\XX) \right\} = \mathbb{E}\left[\Phi\left(\R_{\X}\right)\right],
\end{equation}
where the expectation is with respect to $\X$ (or $\XX$), and
\begin{equation*}
	\Phi(\mathbf{A}):=\tr\left(\cov^{-1}+\frac{T}{\sigma_{\rm s}^{2}}\I_{\Ns}\otimes \mathbf{A}\right)^{-1}
\end{equation*}
is defined for a Hermitian matrix $\mathbf{A}\in\mathbb{C}^{N\times N}$.

\subsubsection{Communication}
The objective of communication for an ISAC system is to convey information reliably for data exchange, therefore, a natural communication metric is the data rate.
For any feasible $p_{\X}$, i.e., those that satisfy the input power constraint \eqref{eqn:power_constraint}, the communication performance (under the assumption of sending a long codeword spanning an infinite number of fading blocks) is given by the ergodic coherent mutual information per unit time
\begin{equation}
	\label{eqn:rate_def}
	\Rate := \frac1T{\mathcal{I}(\X;\Y|\HH)}.
\end{equation}

Under this setup, both $\Error$ and $\Rate$ are functionals of input distribution. Due to the assumption of CSIT, the distribution of $\X$ \emph{can} depend on $\HH$.
In the next section, the optimal input distributions for the MMSE-Rate limit will be investigated.

%% file: limits.tex
\section{Main Theoretical Results}
\label{sec:results}
In this section, the main theoretical results of this paper are presented. Before delving into theoretical details, the ISAC limits are first defined.

\begin{proposition}
	The Pareto-optimal ISAC limit can be obtained by solving the following stochastic functional optimization problem
	\begin{equation} \label{eqn:ISAC_limit}
		\begin{aligned}
			\int \Bigg[\sup_{p_{\X|\HH}\in\mathcal{F}} & \Big( \frac{\alpha}{T} \mathcal{I}(\X;\Y|\HH=\Hr) \\
			&- (1-\alpha)\mathbb{E} \big[ \Phi(\R_{\X})|\HH=\Hr \big]\Big)\Bigg] p_{\HH}(\Hr)\ \diff \Hr,
		\end{aligned}
	\end{equation}
	where $\alpha\in[0,1]$ describes the relative weight of the sensing and communication performance,
	\begin{equation*}
		\mathcal{F} := \Bigg\{p_{\X|\HH}\Big|
		 \mathbb{E}\big\{\tr(\R_{\X})\big|\HH=\Hr\big\} 	\leq N\Po, \forall\Hr\in\mathbb{C}^{\Nc\times N} \Bigg\},
	\end{equation*}
	and $p_{\HH}$ is the distribution of $\HH$.

	\begin{proof}
		Explicitly, the ergodic rate can be written as an integral (expectation) of conditional mutual information for a fixed $\HH = \Hr$: 
        $$\mathcal{I}(\X;\Y|\HH)=\int \mathcal{I}(\X;\Y|\HH=\Hr) p_{\HH}(\Hr)\ \diff \Hr.$$
		According to \cite[Proposition 1]{caire2002capacity}, the relevant maximization problem for the rate with CSIT is
		\begin{equation*}
			\sup_{p_{\X|\HH}(\Xr|\cdot)}
			\mathcal{I}(\X;\Y|\HH)=
			\int \Big[\sup_{p_{\X|\HH}}  \mathcal{I}(\X;\Y|\HH=\Hr) \Big] p_{\HH}(\Hr)\ \diff\Hr.
		\end{equation*}
		From the law of iterated expectation, the average MMSE is
        \begin{equation*}
            \begin{aligned}
                \mathbb{E}\big[\Phi(\R_{\X})\big]&=\mathbb{E}\Big\{\mathbb{E}\big[\Phi(\R_{\X})\big|\HH\big]\Big\}\\
                &=\int \mathbb{E}\big[\Phi(\R_{\X})\big|\HH=\Hr\big]p_{\HH}(\Hr)\ \diff \Hr;
            \end{aligned}
        \end{equation*}
		similar to the mutual information term, the relevant minimization problem for the average MMSE is
		\begin{equation*}
			\sup_{p_{\X}}\ -\mathbb{E}\big[\Phi(\R_{\X})\big]=-\int \Bigg[\sup_{p_{\X|\HH}}\  \mathbb{E}\big[\Phi(\R_{\X})\big|\HH=\Hr\big]\Bigg] p_{\HH}(\Hr)\ \diff\Hr.
		\end{equation*}
		
		It is clear that the conditional mutual information $\mathcal{I}(\X;\Y|\HH=\Hr)$ is concave with respect to $p_{\X|\HH}$ \cite{variation}, and the expected MMSE is linear, thus concave, with respect to $p_{\X|\HH}$. According to the weighted sum scalarization theorem for concave multi-objective problems \cite[Chapter 3]{ehrgott2005multicriteria}, the estimation-information Pareto front can be fully characterized by maximizing a weighted sum of the ergodic rate and the negative average MMSE. As a result, the resulting problem can be expressed as \eqref{eqn:ISAC_limit}, and the support $\mathcal{F}$ is from the short-term power constraint \eqref{eqn:power_constraint}.
	\end{proof}
\end{proposition}

Note that the condition for $\mathcal{F}$ must hold for all $\HH=\Hr$, and the optimizing objective \eqref{eqn:ISAC_limit} is an expectation with respect to $p_{\HH}$ of a function $\alpha T^{-1}\mathcal{I}(\X;\Y|\HH=\Hr)-(1-\alpha)\mathbb{E}[\Phi(\R_{\X})|\HH=\Hr]$ that depends point-wise on each $\HH=\Hr$. Therefore, the solution of \eqref{eqn:ISAC_limit} is found by solving the following point-wise maximization for each $\Hr$, i.e.,
\begin{equation}
	\begin{aligned}
		\sup_{p_{\X|\HH}}\quad& \frac\alpha T \mathcal{I}(\X;\Y|\HH=\Hr)\!-\! (1-\alpha)\mathbb{E}\left[\Phi(\R_{\X})|\HH=\Hr\right] \\
		\rm{s.t.}\quad& \R_{\X}=\R_{\X}^\herm,\ \R_{\X}\succeq\boldsymbol{0},\ 		\mathbb{E}\big\{\tr(\R_{\X})\big|\HH=\Hr\big\} \leq N\Po.
	\end{aligned}
	\label{eqn:ISAC_limit_SAA}
\end{equation}

For any given $\alpha\in[0,1]$ and $\HH=\Hr$, problem \eqref{eqn:ISAC_limit_SAA} is a functional optimization problem. Once the optimal distributions, denoted as $\PXopt$ and $\PYopt=\int\PXopt p_{\Zc}(\Yr-\Hr\Xr)\ \diff\Xr$, are obtained by solving \eqref{eqn:ISAC_limit_SAA}, the sensing and communication performance (conditioned on $\HH=\Hr$) are respectively
$$\begin{aligned}
	\Rate_\alpha\big|(\HH=\Hr) &= \frac1T\Big[h_\alpha(\Y|\HH=\Hr)-h(\Zc)\Big];
	\\
	\Error_\alpha\big|(\HH=\Hr) &= \mathbb{E}\left[\Phi\left(\R_{\X}\right)|\HH=\Hr\right],
\end{aligned}$$
where $$h_\alpha(\Y|\HH=\Hr)=-\int\PYopt\log\PYopt\ \diff \Yr$$ is the conditional differential entropy, and $h(\Zc)$ is the differential entropy of $\Zc$. The overall sensing and communication performances are outer averaging over $\HH$:
\begin{equation*}
		\Rate_\alpha = \mathbb{E} [\Rate_\alpha|\HH];\quad \Error_\alpha = \mathbb{E} [\Error_\alpha|\HH].
\end{equation*}
Finally, the MMSE-Rate performance limit is the set of all MMSE-Rate pairs
\begin{equation}
	\Big\{(\Error_\alpha,\Rate_\alpha)\Big|\alpha\in[0,1]\Big\},
\end{equation}
and the MMSE-Rate performance region is the set of all MMSE-Rate pairs
\begin{equation}
\bigcup_{\alpha\in[0,1]}\Big\{(\Error,\Rate)\Big|0\leq \Rate\leq\Rate_\alpha;\ \Error\geq\Error_\alpha\Big\}.
\end{equation}

Consequently, to obtain the MMSE-Rate performance limit and investigate the operational region of the considered ISAC system, \eqref{eqn:ISAC_limit_SAA} should be solved first.

\subsection{Identifying the MMSE-Rate Performance Limit}
\label{sec:results_limit_identify}
It is first verified that the quantity $\frac{\alpha}{T}
\mathcal{I}(\X;\Y|{\HH=\Hr})$$-(1-\alpha)\mathbb{E}[\Phi(\R_{\X})|{\HH=\Hr}]$ is upper-bounded by the capacity of the fixed communication channel $\Hr$  under power constraint $P_0$ for any $\alpha\in[0,1]$. Hence, the original optimization problem \eqref{eqn:ISAC_limit_SAA} indeed admits a supremum. To identify the ISAC limits, this paper adopts variational methods to solve for the optimal distribution along the entire MMSE-Rate performance limit curve, i.e., $\forall\alpha\in[0,1]$. The necessary conditions for the optimal distribution are given in \textit{Theorem \ref{theorem:limmit_distribution}}.

\begin{theorem}[The Euler-Lagrange Condition for the MMSE-Rate Limit]
	\label{theorem:limmit_distribution}
	Given any communication channel realization $\HH=\Hr$ and $\alpha\in[0,1]$, the following necessary conditions must be satisfied
	\begin{equation} \label{eqn:opt_out_distribution}
		\begin{aligned}
			\alpha\!\int\! p_{\Zc}&(\Yr-\Hr\Xr)  \log \PYopt\ \diff\Yr\\
			 &= T(\alpha-1)\Phi(\Rr_{\Xr})+\mu_{1,\alpha}+\mu_{2,\alpha}\tr(\Xr\Xr^\herm),
		\end{aligned}
	\end{equation}
where $\Rr_{\Xr}:=\dfrac1T \Xr\Xr^\herm$ is the correlation matrix of $\Xr$, $\mu_{1,\alpha}$ and $\mu_{2,\alpha}$ are chosen to guarantee
\begin{subequations}\label{eqn:constraints_condition}
	\begin{align}
		&\int\PXopt\ \diff\Xr=1; \label{eqn:solve_multiplier_1}\\
		&\int\tr(\Xr\Xr^\herm)\PXopt\ \diff\Xr \leq TNP_0. \label{eqn:solve_multiplier_2}
	\end{align}
\end{subequations}

	\begin{proof}
		See \textsc{Appendix \ref{appendix:limit}}.
	\end{proof}
\end{theorem}

\begin{remark}
	Equation \eqref{eqn:opt_out_distribution} is the Euler-Lagrange equation associated with the functional optimization problem \eqref{eqn:ISAC_limit_SAA}. It is a necessary condition for optimality --- any optimal distributions (e.g., $\PXopt$ and $\PYopt$) that actually achieves the MMSE-Rate limits must always satisfy this equation.
\end{remark}

To solve the functional optimization problem, the associated Euler-Lagrange equation \eqref{eqn:opt_out_distribution} should be analyzed. The solution is sought within the \emph{classical function space} $\mathcal{C}$. Specifically, $\mathcal{C}$ consists of all probability density functions (PDFs) that are continuous with respect to the Lebesgue measure and defined on $\mathbb{C}^{N \times T}$. This function space is widely and typically adopted in conventional variational optimization. Additionally, Gaussian inputs, which are optimal for communication systems with Gaussian noise \cite{cover1999elements} and are adopted as a default assumption in many ISAC studies \cite{CRB_BF,CRB_hua}, also belong to \(\mathcal{C}\). Subsequently, distributions belonging to $\mathcal{C}$ are primarily considered. The result is summarized in the following theorem.

\begin{theorem}[Achievability of the MMSE-Rate Limit in $\mathcal{C}$]
\label{theorem:unachivability}
Given any communication channel realization $\HH=\Hr$:
	\begin{itemize}
		\item[$\divideontimes$] For $\alpha=1$, the optimal input distribution $p_{\X|\HH,1}^\star(\Xr|\Hr)\in\mathcal{C}$ and is multi-variate Gaussian.
		\item[$\divideontimes$]
		For $\alpha\in[0,1)$, the MMSE-Rate performance pair $\big\{\Error_\alpha|(\HH=\Hr),\Rate_\alpha|(\HH=\Hr)\big\}$ is unachievable by any PDFs in $\mathcal{C}$.
		\end{itemize}
\begin{proof}
	(Sketch) Both sides of the variational condition \eqref{eqn:opt_out_distribution} are functions of the input variable $\Xr$, and must always hold point-wise for every $\Xr$ over its entire support. To verify, rank-1 matrices are used as a counterexample, which form a proper subset of the full support. Via the Hermite transform, the left-hand side of \eqref{eqn:opt_out_distribution} is proven to be a polynomial of $\Xr$ defined globally. In contrast, due to the structure of $\Phi(\cdot)$, the right-hand side is a polynomial of $\Xr$ only within a limited region. For $\alpha \in [0,1)$, the condition \eqref{eqn:opt_out_distribution} fails to hold for all rank-1 inputs. This indicates that general PDFs in $\mathcal{C}$ with full support cannot satisfy the necessary condition for optimality. For detailed proofs, please see Appendix \ref{appendix:unachievability}. 
\end{proof}
\end{theorem}

\begin{remark}[Structure of the Communication-Optimal Input Distribution]
	It is well known that the communication-optimal (e.g., when $\alpha=1$) waveform is Gaussian. In Theorem \ref{theorem:unachivability}, this result is obtained directly through a variational analysis of the objective functional, rather than by invoking the prior knowledge that ``Gaussian inputs maximize mutual information under Gaussian channels''.
\end{remark}

\begin{remark} \label{Remark:unachievability}
	For general points on the MMSE-Rate curve except for $\Pc$ (e.g., $0\leq \alpha<1$), the limit is generally unachievable by any PDF in $\mathcal{C}$, as no valid PDF in $\mathcal{C}$ satisfies the necessary condition specified in Theorem \ref{theorem:limmit_distribution}. Consequently, the optimum of problem \eqref{eqn:ISAC_limit_SAA} is a supremum in $\mathcal{C}$, not a maximum. The optimal (capacity-achieving) input distribution must lie outside the set $\mathcal{C}$, and can only be:
		\begin{itemize}
    		\item[i)] discrete, singular, or a mixture thereof; and/or
    		\item[ii)] supported on a proper subset of $\mathbb{C}^{N \times T}$ determined by additional constraints (e.g., rank).
		\end{itemize}
	This implies that common continuous distributions whose realizations can take arbitrary values in $\mathbb{C}^{N\times T}$ (e.g., Gaussian, Laplace, Cauchy, etc.), are excluded from optimal ISAC waveforms under these conditions.
\end{remark}

\begin{corollary}
	\label{corollary:S-opt-correlation}
	When $\alpha=0$, the optimal input distribution $p_{\X|\HH,0}^\star(\Xr|\Hr)$ is required to preserve a deterministic sample correlation matrix, e.g., $\R_{\X}=\tilde{\Rr}_{\X}$. More importantly, $p_{\X|\HH,0}^\star(\Xr|\Hr)$ is singular and supported by a restricted subset of $\mathbb{C}^{N\times T}$, and $p_{\X|\HH,0}^\star(\Xr|\Hr)\notin \mathcal{C}$.
	\begin{proof}
	When $\alpha = 0$, equation \eqref{eqn:opt_out_distribution} reduces to
 		$$-T \Phi(\Rr_{\Xr}) + \mu_{1,0} + \mu_{2,0}T\tr(\Rr_{\Xr}) = 0.$$

 	It can be shown that $$\frac{\partial}{\partial \tr(\Rr_{\Xr})}\left[-T \Phi(\Rr_{\Xr}) + \mu_{1,0} + \mu_{2,0} T\tr(\Rr_{\Xr})\right]<0.$$ Hence, the above equation has a unique solution with respect to $\tr(\Rr_{\Xr})$, and the solution is $\tr(\Rr_{\Xr})=NP_0$. Consequently, $\Phi(\Rr_{\Xr})=\mu_{2,0}NP_0+\mu_{1,0}/T$ is a constant. It is straightforward to verify that $\Phi(\Rr_{\Xr})$ indeed is the minimized average MMSE. Note that $\Phi(\cdot)$ is a strictly convex function. According to the Jensen's inequality $\mathbb{E}[\Phi(\R_{\X})]\geq \Phi(\mathbb{E}[\R_{\X}])=\Phi(\tilde{\Rr}_{\X})$, the minimized average MMSE is $\min_{\tilde{\Rr}_{\X}} \Phi(\tilde{\Rr}_{\X})$, corresponding to a unique $\tilde{\Rr}_{\X}$. Therefore, $\Phi(\Rr_{\Xr})=\min_{\tilde{\Rr}_{\X}}\Phi(\tilde{\Rr}_{\X})$ implies that every valid realization of $\X$ (i.e., those within the support of $p_{\X|\HH,0}^\star$) must yield a sample correlation matrix $\R_{\X}$ equal to this unique $\tilde{\Rr}_{\X}$ that minimizes the average MMSE. In other words, the sample correlation matrix must be deterministic, hence equal to the statistical correlation matrix, and both must satisfy the above equation.
	\end{proof}
\end{corollary}

\begin{remark}[Structure of the Sensing-Optimal Input Distribution]
	For the sensing-optimal waveform (e.g., when $\alpha = 0$), although $\X$ must be random to provide a certain level of entropy, its sample correlation matrix $\R_{\X}$ must be deterministic in order to optimize sensing performance. This observation is well-aligned with the CRB-minimal random waveform discussed in \cite{FanLiu}, wherein the trace of the sample correlation matrix is required to be fixed. These observations reflect a common preference across different sensing metrics: sensing generally favors signals with lower randomness to ensure performance.
\end{remark}

When the input distribution is not full-support, further applying variational methods to characterize the optimal distribution becomes intractable. This is because the optimal solution and its support are inherently coupled to satisfy the variational condition \eqref{eqn:opt_out_distribution}. To evaluate and approach the limit, one must construct a sequence of probability density functions in $\mathcal{C}$ that asymptotically satisfy the variational conditions and yield progressively improved performance. This requirement aligns naturally with the iterative structure of the B-A algorithm \cite{Blahut,Arimoto}, which successively refines input distributions to approach optimality. Motivated by this connection, this paper proceeds to numerically investigate the achievable MMSE-Rate performance and limit-approaching strategies using a B-A type algorithm.

\begin{algorithm*}[!t]
	\caption{Constrained Blahut-Arimoto Type Algorithm for Evaluating the MMSE-Rate Limit}
	\label{algorithm:BA}
	\KwIn{$\alpha$, $\Hr$, $\varepsilon_{\!J}$, $\varepsilon_{\mu}$}
	\KwOut{$\PXopt$, $\Rate_\alpha|(\HH=\Hr)$, and $\Error_\alpha|(\HH=\Hr)$}
	\Indp
	\BlankLine
	\textbf{Initialize:}\\
	(1) Calculate $p_{\Y|\X,\HH}(\Yr|\Xr,\Hr)=p_{\Zc}(\Yr-\Hr\Xr)$\\
	(2) Choose $p_{\X|\HH,\alpha}^{(0)}(\Xr|\Hr)$ such that $p_{\X|\HH,\alpha}^{(0)}(\Xr|\Hr)\geq0$ and $\int p_{\X|\HH,\alpha}^{(0)}(\Xr|\Hr)\ \diff\Xr=1$\\
	(3) Calculate
	$$\begin{aligned}
		D^{(0)}_\alpha(\Hr) &= \int p_{\Y|\X,\HH}(\Yr|\Xr,\Hr)\log\frac{p_{\Y|\X,\HH}(\Yr|\Xr,\Hr)}{\int p_{\X|\HH,\alpha}^{(0)}(\Xr|\Hr) p_{\Y|\X,\HH}(\Yr|\Xr,\Hr)\ \diff\Xr}\ \diff \Yr \\
		J^{(0)}_\alpha(\Hr) &= \int p_{\X|\HH,\alpha}^{(0)}(\Xr|\Hr)\left[\frac{\alpha}{T}D^{(0)}_\alpha(\Hr)-(1-\alpha) \Phi(\Rr_{\Xr})\right]\ \diff\Xr
	\end{aligned}$$\\
	(4) $i=0$\\
	(5) $\mu^{(0)}=-1$\\
	
	\Repeat{$|J^{(i)}_\alpha(\Hr)-J^{(i-1)}_\alpha(\Hr)|\leq \varepsilon_{\!J}$}{
		$i=i+1$ \\
		\textbf{Initialize:}\\
		(1) $j=0$\\
		(2) $\mu^{(i,0)}=\mu^{(i-1)}$\\
		\Repeat{$|\mu^{(i,j)}-\mu^{(i,j-1)}|\leq\varepsilon_{\mu}$}{
			$j=j+1$ \\
			Calculate
			$$\begin{aligned}
				\mu&^{(i,j)} = \mu^{(i,j-1)}\\
				&-\frac{
					\int \left(1-\frac{\tr(\Xr\Xr^\herm)}{NT\Po}\right){p_{\X|\HH,\alpha}^{(i-1)}(\Xr|\Hr)}\exp\left[D^{(i-1)}_\alpha(\Hr)+\mu^{(i,j-1)}\tr(\Xr\Xr^\herm)-(\frac1\alpha-1)T\Phi(\Rr_{\Xr})\right]\ \diff\Xr}{\int \tr(\Xr\Xr^\herm)\left(1-\frac{\tr(\Xr\Xr^\herm)}{NT\Po}\right){p_{\X|\HH,\alpha}^{(i-1)}(\Xr|\Hr)}\exp\left[D^{(i-1)}_\alpha(\Hr)+\mu^{(i,j-1)}\tr(\Xr\Xr^\herm)-(\frac1\alpha-1)T\Phi(\Rr_{\Xr})\right]\ \diff\Xr}
			\end{aligned}$$\\
			\If{$\mu^{(i,j)}>0$}{
				$\mu^{(i,j)}=0$
			}
		}	
		$\mu^{(i)}=\mu^{(i,j)}$ \\
		Calculate
		\begin{equation*}
			\begin{aligned}
				p_{\X|\HH,\alpha}^{(i)}(\Xr|\Hr) &= \frac{p_{\X|\HH,\alpha}^{(i-1)}(\Xr|\Hr)\exp D^{(i-1)}_\alpha(\Hr)\exp\left[\mu^{(i)}\tr(\Xr\Xr^\herm)-(\frac1\alpha-1)T\Phi(\Rr_{\Xr})\right]}{\int p_{\X|\HH,\alpha}^{(i-1)}(\Xr|\Hr)\exp D^{(i-1)}_\alpha(\Hr)\exp\left[\mu^{(i)}\tr(\Xr\Xr^\herm)-(\frac1\alpha-1)T\Phi(\Rr_{\Xr})\right]\ \diff\Xr}\\
				D^{(i)}_\alpha(\Hr) &= \int p_{\Y|\X,\HH}(\Yr|\Xr,\Hr)\log\frac{p_{\Y|\X,\HH}(\Yr|\Xr,\Hr)}{\int p_{\X|\HH,\alpha}^{(i)}(\Xr|\Hr) p_{\Y|\X,\HH}(\Yr|\Xr,\Hr)\ \diff\Xr}\ \diff \Yr
			\end{aligned}
		\end{equation*} \\
		Calculate
		$$\begin{aligned}
			R^{(i)}_\alpha(\Hr)&=\frac1T\int p_{\X|\HH,\alpha}^{(i)}(\Xr|\Hr)D^{(i)}_\alpha(\Hr)\ \diff\Xr\\
			E_\alpha^{(i)}(\Hr)&=\int p_{\X|\HH,\alpha}^{(i)}(\Xr|\Hr)\Phi(\Rr_{\Xr})\ \diff\Xr\\
			J^{(i)}_\alpha(\Hr) &= \alpha R^{(i)}_\alpha(\Hr)-(1-\alpha)E^{(i)}_\alpha(\Hr)
		\end{aligned}$$
	}
	
	{\bf Return:} $\PXopt=p_{\X|\HH,\alpha}^{(i)}(\Xr|\Hr)$, $\Rate_\alpha|(\HH=\Hr)=R^{(i)}_\alpha(\Hr)$, and $\Error_\alpha|(\HH=\Hr)=E^{(i)}_\alpha(\Hr)$
\end{algorithm*}

\subsection{Computing The MMSE-Rate Performance Limit: Blahut-Arimoto Type Algorithm}
\label{sec:results_limit_compute}

The B-A algorithm \cite{Blahut,Arimoto} is a classical numerical method originally designed for computing the capacity of discrete memoryless channels (DMCs). It operates based on alternating optimization: by iteratively updating the input distribution and the posterior transition probability, the algorithm converges to an optimal input that maximizes mutual information. This core principle of the B-A algorithm has been widely adopted and extended to calculate the capacity and capacity-achieving distribution under various conditions or constraints, including solving lossy source coding rate-distortion function \cite{Blahut}, information bottleneck analysis \cite{BA_information_bottleneck}, etc. In \cite{ISAC_State}, the B-A algorithm is adapted for a general sensing distortion function for ISAC with DMC, where the focus is on coding strategies.

Next, this paper develops a B-A type algorithm customized for evaluating the MMSE–Rate limits and the corresponding limit-converging input distributions. To this end, several essential points are highlighted to adapt the classical B-A algorithm to the specific problem formulation in this paper:
\begin{itemize}
	\item[i)] an additional MMSE-related term, e.g., the term $(1-\alpha)\mathbb{E}\left[\Phi(\R_{\X})|\HH=\Hr\right]$, in the objective function \eqref{eqn:ISAC_limit_SAA} is included, which captures the channel estimation performance;
	\item[ii)] the channels are continuous-input/output in contrast to the conventional B-A algorithm for DMC. Hence, summation over discrete input and output symbols in the classical B-A algorithm is replaced by integration over continuous-valued, high-dimensional matrices; and
	\item[iii)] a numerical procedure for solving Lagrange multipliers appears as part of the proposed algorithm;
	\item[iv)] a proof of convergence of the proposed algorithm to the MMSE-Rate limit is provided.
\end{itemize}
These adaptations ensure that the proposed algorithm remains consistent with the specific framework and setup in this paper while preserving the spirit of the original B-A procedure.

Given $\alpha$ and a specific channel realization $\Hr$, the procedure for evaluating the MMSE-Rate limit is summarized in Algorithm~\ref{algorithm:BA}. Specifically, the input parameters $\varepsilon_{\!J}$ and $\varepsilon_{\mu}$ specify the numerical tolerances for the evaluation accuracy of the MMSE-Rate limit and the convergence of the Lagrange multipliers, respectively. The derivation and proof of convergence of Algorithm \ref{algorithm:BA} can be found in \textsc{Appendix \ref{appendix:BA}}.

\begin{remark}[Computational Complexity of the Proposed Algorithm]
	The proposed B-A type algorithm can be computationally expensive in high-dimensional settings, as numerical integration over $\Xr$ may scale exponentially with the dimension of $\Xr$ and $\Yr$ when using standard methods (e.g., Monte Carlo). To evaluate the ergodic rate, evaluations of the algorithm for different $\Hr$ will be further required to take the expectation.
	To ensure tractability, this paper restricts numerical studies to low-dimensional cases (see details in Section \ref{sec:numerical}). This complexity arises from the intrinsic nature of the iterative integration steps in the B-A algorithm itself, rather than any limitation of the underlying MMSE–Rate formulation. Developing scalable variants (e.g., using structural approximations, variational inference, neural network approximation, etc.) remains an important issue subject to future research.
\end{remark}

%% file: SC_opt_point_result.tex
\subsection{Sensing- and Communication-Optimal Performance Characterization and Discussion}
\label{sec:results_SAC}
Sensing- and communication-optimal points are specifically investigated in this part. There are several main reasons:
\begin{itemize}
	\item[i)] The performance at $\Ps$ and $\Pc$ are critical for designing sensing- and communication-centric ISAC systems, as well as for gaining deeper insights into the sensing-communication tradeoff \cite{CRB_BF,CRB_hua,FanLiu};
	\item[ii)] Characterizations of $\Ps$ and $\Pc$ under the MMSE-ergodic rate framework have not been reported and do not follow as a simple extension of existing results;\footnote{Although \cite{FanLiu} studies the two optimal points under the CRB-Rate framework, their analysis is focused on structural and qualitative insights. In contrast, this work focuses on the MMSE-ergodic rate and provides explicit \emph{closed-form} characterizations.}
	\item[iii)] According to \emph{Theorem \ref{theorem:unachivability}}, only $\Ps$ and $\Pc$ admit relatively clear achieving strategies (although $\Ps$ cannot be attained by any valid distribution in $\mathcal{C}$);
	\item[iv)] The performance and achieving strategy at $\Ps$ (e.g., $\alpha=0$) cannot be numerically obtained from the B-A algorithm, thus requiring explicit investigation.
\end{itemize}
This paper proceeds to provide a detailed investigation of these two optimal points under the MMSE-Rate criterion. Specifically, \emph{closed-form} optimal waveforms (and optimal estimators) and corresponding ISAC performance are derived. The MMSE-Rate tradeoff is explicitly characterized under these waveforms, revealing fundamental insights between sensing and communication. Furthermore, loose bounds on the CRB-Rate framework \cite{FanLiu} are revisited and derived to accommodate the MMSE-Rate setup.

\subsubsection{Sensing and Communication Optimal Waveform}

\paragraph{Sensing-Optimal Point and Its ISAC Performance}

The sensing-optimal point $\Ps(\Es,\Rate_{\rm s})$ characterizes the best communication performance limited by the ISAC sensing-only scenario, where the channel estimation MSE should be minimized. With $\alpha=0$, \eqref{eqn:ISAC_limit} reduces to
\begin{equation*}
	\int\!\! \sup_{p_{\X|\HH}\in\mathcal{F}} \Big(-\mathbb{E} \big[ \Phi(\R_{\X})|\HH=\Hr \big]\Big)p_{\HH}(\Hr)\ \diff \Hr=\inf_{p_{\X}\in\mathcal{F}}\ \mathbb{E}[\Phi(\R_{\X})],
\end{equation*}
where the dependency on $\HH$ of $p_{\X|\HH}$ disappears as the sensing channel is independent of $\HH$. The best sensing performance $\Es$ is thus
\begin{equation}
	\begin{aligned} 
		\Es= &\inf_{p_{\X}}& &\mathbb{E}\left[\Phi\left(\R_{\X}\right)\right]
		\\
		&\rm{s.t.}& &\R_{\X}=\R_{\X}^\herm,\ \R_{\X}\succeq\mathbf{0},\ \mathbb{E}\{{\tr}(\R_{\X})\}=N\Po.
	\end{aligned}
	\label{eqn:sensing_opt}
\end{equation}

Denote $\X_{\rm s}$ as the MMSE-minimal waveform that achieves $\Es$. In \textit{Theorem \ref{theorem:sensing_opt}}, it is shown that the distribution of $\X_{\rm s}$ is not unique. Therefore, the sensing-limited communication performance is defined as the maximum achievable rate over all distributions in the set of MMSE-minimal waveforms, i.e.,
\begin{equation}
	\begin{aligned}
		\Rate_{\rm s} = \sup_{p_{\X_{\rm s}}}\quad T^{-1} \mathcal{I}(\X_{\rm s};\Y|\HH),
	\end{aligned}
\end{equation}
where $p_{\X_{\rm s}}$ is the solution that optimizes \eqref{eqn:sensing_opt}.

In \textit{Corollary \ref{corollary:S-opt-correlation}}, it is proven from the variational condition that the sensing-optimal sample correlation matrix (if it exists) must be unique and deterministic, e.g., $\R_{\X_{\rm s}}=\tilde{\Rr}_{\X_{\rm s}}$. Indeed, the sensing-optimal waveform still maintains randomness, but due to the deterministic requirements on its sample correlation matrix, its randomness is greatly reduced. In the next theorem, this paper derives the sensing-optimal waveform $\X_{\rm s}$ and its corresponding MMSE $\Es$.

\begin{theorem}[Sensing-Optimal Waveform and its Corresponding MMSE] \label{theorem:sensing_opt}
	The sensing-optimal waveform is\footnote{This paper refers to the sensing-optimal signal as ``isometry signal'' hereafter, since $\stiefel$ preserves the length of any vector, i.e., $\left\|\stiefel^\herm\boldsymbol{x}\right\|=\left\|\boldsymbol{x}\right\|$, $\forall\boldsymbol{x}\in\mathbb{C}^{N\times 1}$.}
	\begin{equation}
		\X_{\rm s}=\Ut\powers^{1/2}\stiefel.
		\label{eqn:sensing_opt_waveform}
	\end{equation}
	Particularly, $\Ut$ is from the decomposition of $\covt$ in \eqref{eqn:eigen_decomosition_G},
	\begin{equation}
		\powers=\sigma_{\rm s}^2\diag\left[\varrho_1^+(\ell),\varrho_2^+(\ell),\cdots,\varrho_N^+(\ell)\right],
		\label{eqn:sensing_opt_power}
	\end{equation}
	where $\varrho_i(\ell)$ satisfies $\varphi_i\Big(\varrho_i(\ell),\ell\Big)=0$ with
	\begin{equation}
		\varphi_i(\varrho,\ell) := \sum_{j=1}^{\Ns}(\sigma_{\mathsf g}^{-2}\eigr{j}^{-1}\eigt{i}^{-1}+\varrho)^{-2}-\ell,
		\label{eqn:waterfilling_equation}
	\end{equation}
	and $\ell$ is chosen to guarantee $\tr(\powers)=TNP_0$, and ${\stiefel}\in\mathbb{C}^{N\times T}$ is any semi-unitary matrix, corresponding to a horizontal slice of $N$ rows of a unitary matrix in $\mathbb{C}^{T\times T}$, i.e., $\stiefel\stiefel^\herm=\I_{N}$. The sensing-optimal MMSE is
	\begin{equation}
		\Es = \sum_{\substack{1 \le i \le N \\ 1 \le j \le \Ns}}\frac{\sigma_{\mathsf g}^2\eigr{j}\eigt{i}}{1+\sigma_{\mathsf g}^2\eigr{j}\eigt{i}\varrho^+_i(\ell)}.
		\label{eqn:sensing_opt_MMSE}
	\end{equation}
	Furthermore, the sensing-optimal correlation matrices are
	\begin{equation}
		\R_{\X_{\rm s}}=\tilde{\Rr}_{\X_{\rm s}}=\frac1T\Ut\powers\Ut^\herm.
		\label{eqn:sensing_opt_correlation_analytical}
	\end{equation}
	\begin{proof}
	See \textsc{Appendix} \ref{proof:sensing_opt}.
\end{proof}
\end{theorem}

\begin{remark}
	From Theorem \ref{theorem:sensing_opt}, it can be seen that the sensing-optimal waveform $\X_{\rm s}$
	\begin{itemize}
		\item[i)] depends solely on the eigenvector space of the Tx-side correlation matrix of the sensing channel and is independent of that of the Rx-side;
		\item[ii)] employs the water-filling-like\footnote{It should be noted that \eqref{eqn:waterfilling_equation} is significantly different from the conventional water-filling solution and $\powers$ does not admit a closed-form expression in general.} power allocation strategy, where more power is assigned to subchannels with larger eigenvalues of the Tx-side correlation matrix; and
		\item[iii)] has randomness only in the matrix $\stiefel$ that can be used to modulate information, as $\Ut$ and $\powers$ are completely deterministic and determined by the sensing channel statistics.
	\end{itemize}
\end{remark}

\begin{corollary}
	\label{corollary:sensing_opt_reduced}
	For the special case when antennas at sensing Rx are uncorrelated (e.g., the receiving antennas are widely-separated \cite{sensing_channel_assumption1,MIMO-Radar-waveform,Sensing_channel_model,Widely-Separated-Antennas}), the Rx-side correlation $\covr=\I_{\Ns}$, the sensing-optimal waveform employs water-filling power allocation strategy, and $\powers$ in \eqref{eqn:sensing_opt_power} admits closed-form expression given as
	\begin{equation}
		\powers=\diag\left[\Big(\eta_{\rm s}-\frac{\sigma_{\rm s}^2}{\sigma_{\mathsf g}^2 \eigt{1}}\Big)^{\!+}\!\!\!,\cdots,\Big(\eta_{\rm s}-\frac{\sigma_{\rm s}^2}{\sigma_{\mathsf g}^2 \eigt{N}}\Big)^{\!+}\right],
		\label{eqn:sensing_opt_power_reduced}
	\end{equation}
	 where $\eta_{\rm s}$ is chosen such that
	\begin{equation}
		\sum_{i=1}^{N}\left({\eta}_{\rm s}-\frac{\sigma_{\rm s}^2}{\sigma_{\mathsf g}^2\eigt{i}}\right)^+ = TN\Po,
		\label{eqn:sensing_opt_sum_power_reduced}
	\end{equation}
	the resulting MMSE \eqref{eqn:sensing_opt_MMSE} reduces to
	\begin{equation}
		\Es=\Ns  \sum_{i=1}^{N} \min\Big\{\eigt{i}\sigma_{\mathsf g}^2,\eta_{\rm s}^{-1}\sigma_{\rm s}^2\Big\}.
		\label{eqn:sensing_opt_MMSE_reduced}
	\end{equation}
	The best MMSE estimator for the sensing channel that achieves \eqref{eqn:sensing_opt_MMSE_reduced} can be derived as
	\begin{equation}
		\hat{\g} =\I_{\Ns}\otimes\left[\Ut \diag(\tilde{p}_1,\tilde{p}_2,\cdots,\tilde{p}_N)\stiefel \right]\s,
		\label{eqn:est_results}
	\end{equation}
	where $\tilde{p}_i:=\dfrac{1}{\eta_{\rm s}}\sqrt{\left(\eta_{\rm s}-\frac{\sigma_{\rm s}^2}{\sigma_{\mathsf g}^2\eigt{i}}\right)^+}$. Accordingly,
	\begin{equation}
		\hat{\G} = \Sc\stiefel^\herm \diag(\tilde{p}_1,\tilde{p}_2,\cdots,\tilde{p}_N)\Ut^\herm.
	\end{equation}
	\vspace{0.15cm}
	\begin{proof}
		See Appendix \ref{proof:sensing_opt_reduced}.
	\end{proof}
\end{corollary}

\begin{remark}
	As shown in \eqref{eqn:sensing_opt_power_reduced}, the sensing-optimal power allocation employs the water-filling strategy when Rx-side sensing channels are uncorrelated. It allocates the transmitted power in reverse proportion to the eigenvalue of $\covt$. An intuitive understanding is that water-filling focuses the signal power in the direction with the presence of significant scattering of the target, and allocates less power in the direction with less prominent target scattering compared with the noise, i.e., where $\sigma_{\rm s}^2/(\sigma_{\mathsf g}^2\eigt{i})$ is large.
\end{remark}

 To investigate the best communication performance, i.e., the maximum rate, achieved by $\X_{\rm s}$ is typically intractable. This is because the sensing-optimal distribution is not well-defined in $\mathcal{C}$, e.g., it is singular with restricted support, thus a variational approach discussed in Section \ref{sec:results_limit_identify} fails to apply under this specific scenario. Thanks to \cite{FanLiu,Grassmann}, the sensing-limited capacity is proven to be achieved when $\stiefel$ is uniformly sampled from the set of all orthonormal matrices, i.e., the complex Stiefel manifold $\{\stiefel\in\mathbb{C}^{N\times T}|\stiefel\stiefel^\herm={\I}_N\}$.

\begin{theorem}[Sensing-Limited High Signal-to-Noise Ratio Ergodic Rate] \label{theorem:Ps_rate_high_SNR}
	In the high SNR region (i.e., when $\Po/\sigma_{\rm c}^2$ very large), the rate can be expressed as
\begin{equation}
	\begin{aligned}
			\Rate_{\rm s} &= \mathbb{E}\left[
		\left(1-\frac{\Ms(\HH)}{2T}\right)\log\det\left(\frac{\HH\Ut\powers\Ut^\herm\HH^\herm}{\sigma_{\rm c}^2T}\right)+c_0(\HH)\right]\\
		&\hspace{1cm} +\mathcal{O}(\sigma_{\rm c}^2),
	\end{aligned}
	\label{eqn:sensing-limited_rate}
\end{equation}
where $\Ms(\HH)=\rank(\HH\Ut\powers\Ut^\herm\HH^\herm)$, and
\begin{equation*}
	c_0(\HH)=\frac{\Ms(\HH)}{T}\left[\left(T-\frac{\Ms(\HH)}2\right)\log\frac{T}{\e}-\log\Gamma(T)+\log(2\sqrt\pi)\right],
\end{equation*}
with $\e$ being the Euler's constant, and $\Gamma(\cdot)$ is the Gamma function.
	\begin{proof}
		See \cite{Grassmann} and \cite[\textit{Theorem 1}]{FanLiu} and replace the sensing-optimal statistical correlation matrix with \eqref{eqn:sensing_opt_correlation_analytical}. 
	\end{proof}
\end{theorem}

\paragraph{Communication-Optimal Waveform and Its ISAC Performance}

The communication-optimal point $\Pc(\Ec,\Rc)$ characterizes the best sensing performance limited by the ISAC communication-only scenario, where the rate should be maximized. The best communication performance $\Rate_{\rm c}$ is obtained by solving \eqref{eqn:ISAC_limit} with $\alpha=1$:
\begin{equation}
	\int \sup_{p_{\X|\HH}\in\mathcal{F}} \Big(\frac{1}{T} \mathcal{I}(\X;\Y|\HH=\Hr) \Big)p_{\HH}(\Hr)\ \diff \Hr.
	\label{eqn:rate_opt_problem}
\end{equation}
To find the ergodic maximal rate (capacity), it suffices to research on one coherence time $T$, corresponding to one channel realization $\Hr$ \cite{Grassmann,caire2002capacity}.

\begin{theorem}[Closed-Form Communication-Optimal Waveform and Its Corresponding ISAC Performance]
\label{theorem:comm_opt}
	For a given channel realization $\Hr$ in each block, the communication-optimal waveform is
	\begin{equation}
		\begin{aligned}
			\Big(\X_{\rm c}|\HH=\Hr\Big)&=\UH\powerc^{1/2}(\Hr)\N\\
			&\sim\mathcal{MCN}(\mathbf{0}_{N\times T},\UH\powerc(\Hr)\UH^\herm,\I_{T}),
		\end{aligned}
		\label{eqn:comm_opt_waveform}
	\end{equation}
	where $\UH\in\mathbb{C}^{N\times N}$ is from the decomposition $\Hr^\herm\Hr=\UH\boldsymbol{\mathit\Gamma}_{\!\Hr}\UH^\herm$ with $r_{\Hr}=\rank(\Hr)$,
	\begin{equation}
		\begin{aligned}
			\powerc(\Hr)=\diag\Bigg[\bigg(\eta_{\rm c}(\Hr&)-\frac{\sigma_{\rm c}^2}{\gamma_1(\Hr)}\bigg)^+,\cdots,\\
			&\left(\eta_{\rm c}(\Hr)-\frac{\sigma_{\rm c}^2}{\gamma_{r_{\Hr}}(\Hr)}\right)^+,\underbrace{0,\cdots,0}_{N-r_{\Hr}}\Bigg],
		\end{aligned}
		\label{eqn:comm_opt_power}
	\end{equation}
	with $\gamma_i(\Hr)=[\boldsymbol{\mathit\Gamma}_{\!\Hr}]_{ii}$, $\eta_{\rm c}(\Hr)$ being chosen such that
	\begin{equation}
		\sum_{i=1}^{r_{\Hr}}\left(\eta_{\rm c}(\Hr)-\frac{\sigma_{\rm c}^2}{\gamma_i(\Hr)}\right)^+=N\Po,
		\label{eqn:com_opt_power_calculation}
	\end{equation}
	and $\N\in\mathbb{C}^{N\times T}$ has i.i.d. zero-mean circular symmetrical complex Gaussian entries with unit variance, i.e., $\mathbb{E}(\N\N^\herm)=T\I_{N}$. The ergodic coherent rate at $\Pc$ is
	\begin{equation}
			\Rate_{\rm c}=\ \mathbb{E}\left[\sum_{i=1}^{r_{\HH}}\Big[\log\big(\sigma_{\rm c}^{-2}\eta_{\rm c}(\HH)\gamma_i(\HH)\big)\Big]^+\right];
		\label{eqn:comm_optimal_rate}
	\end{equation}
	the channel estimation performance is
	\begin{equation}
		\Ec = \mathbb{E}\Big\{\mathbb{E}\left[
		\Phi\left(\R_{\X}\right)\big|\HH\right] \Big\},
	\label{eqn:com_opt_error}
\end{equation}
where the inner expectation is with respect to $\X_{\rm c}|\HH$, while the outer expectation is with respect to $\HH$.
\end{theorem}

\begin{remark}
	Unlike the sensing-optimal waveform $\X_{\rm s}$, where $\stiefel$ only needs to maintain orthogonality and its distribution still has room for optimization to modulate data, the distribution of the communication-optimal waveform is fully determined --- it is completely specified by the given communication channel realization $\Hr$.
\end{remark}

%% file: SC_opt_point_discussion.tex
\subsubsection{Tradeoffs between Sensing and Communication under the MMSE-Rate Framework}

Reviewing the limit-achieving conditions (Section \ref{sec:results_limit_identify}) and structure of closed-form optimal waveforms (Section \ref{sec:results_SAC}) derived so far, this paper highlights the presence of two tradeoffs between sensing and communication, which are summarized below.
\paragraph{Water-Filling Tradeoff} 

It can be seen that both $\X_{\rm s}$ and $\X_{\rm c}|\HH$ employ a water-filling(-like) power allocation strategy (to the eigenspace of sensing and communication channels, respectively). Indeed, the total ISAC signal power is shared between sensing and communication, but it's not possible to simultaneously achieve the optimal water-fillings for both sensing and communication. Thus, a water-filling tradeoff (WFT) naturally exists, which can be controlled by allocating the power of the transmitted signal more ``aligned'' in favor of either the sensing or communication channel. This WFT provides a refined and formal characterization of the “subspace tradeoff” for the CRB-rate framework proposed in \cite{FanLiu}, under a tractable MMSE-rate framework.

\paragraph{Waveform Uncertainty Tradeoff} 

While sensing prefers predictable waveforms \cite{DRT}, purely deterministic signals cannot convey information. ISAC thus requires waveforms with controlled randomness to balance both functions. Beyond sensing- and communication-optimal distributions, the variational condition derived in \eqref{eqn:opt_out_distribution} reveals the overall trend along the performance limit: when $\alpha \to 1$ (communication-dominant), the influence of the estimation error term on the optimal distribution diminishes, leading to a more Gaussian-like distribution; conversely, when $\alpha \to 0$ (sensing-dominant), the effect of the Gaussian convolution term weakens, and the optimal distribution shifts towards isometry (as verified in Figs. \ref{fig:SISO_distributions} and \ref{fig:MIMO_distribution}). This paper highlights the inherent waveform uncertainty tradeoff (WUT) in ISAC, which can be adjusted via waveform design (e.g., Gaussian, isometry, or other structured random signals from the B-A type algorithm). This insight aligns with prior ISAC studies \cite{FanLiu,Guo,DRT}, despite differences in specific metrics. In \cite{FanLiu}, this fact has been identified under the name of ``deterministic-random tradeoff'' from the sensing- and communication-optimal points, i.e., the two end points of the Pareto boundary. In contrast, our analysis is established for the entire Pareto curve using variational analysis, thereby providing a more general insight.

\subsubsection{Loose Bounds Connecting $\Ps$ and $\Pc$} 

The evaluation of the MMSE-Rate performance limit-converging distribution based on Algorithm \ref{algorithm:BA} is computationally intensive. For fast evaluation of the MMSE-Rate limit, loose bounds may be obtained by leveraging the above-mentioned two tradeoffs.

\paragraph{MMSE-Rate Outer Bound} A relaxed version of \eqref{eqn:ISAC_limit} may be considered for an outer bound. For any given communication channel realization $\Hr$, following inequalities hold
\begin{subequations} \label{eqn:outer_bound_relax}
	\begin{align} 
		&T^{-1}\mathcal{I}(\X;\Y|\HH=\Hr) \leq \log\det\left(\I+\frac{\Hr\tilde{\Rr}_{\X}\Hr^\herm}{\sigma_{\rm c}^2}\right); \label{eqn:outer_bound_relax_com}\\
		&\mathbb{E}\left[\Phi\left(\R_{\X}\right)|\HH=\Hr\right] \geq \Phi\Big(\mathbb{E}[\R_{\X}]\big|\HH=\Hr\Big)=\Phi\Big(\tilde{\Rr}_{\X}\Big), \label{eqn:outer_bound_relax_sen}
	\end{align}
\end{subequations}
a natural loose (outer) bound can be obtained \cite{CRB_hua,CRB_BF,FanLiu} from the following deterministic optimization problem
\begin{equation}
	\begin{aligned}
		\sup_{\tilde{\Rr}_{\X}}\quad& \alpha\log\det\left(\I+\frac{\Hr\tilde{\Rr}_{\X}\Hr^\herm}{\sigma_{\rm c}^2}\right) - (1-\alpha) \Phi\Big(\tilde{\Rr}_{\X}\Big) \\
		\rm{s.t.}\quad& \tilde{\Rr}_{\X}=\mathbb{E}\Big[\frac1T\X\X^\herm\Big],\ {\tr}\Big(\tilde{\Rr}_{\X}\Big) = N\Po.
	\end{aligned} \label{eqn:ISAC_outer_bound}
\end{equation}
Problem \eqref{eqn:ISAC_outer_bound} optimizes with respect to the second-order statistics, e.g., the correlation matrix $\tilde{\Rr}_{\X}$, which constitutes a standard concave optimization problem. Similar formulations have been thoroughly investigated in prior literature \cite{CRB_hua,CRB_BF}, and detailed derivations are omitted here due to space limitations. For any given $\alpha\in[0,1]$, denote the optimal statistical correlation matrix to \eqref{eqn:ISAC_outer_bound} as $\tilde{\Rr}_{\X,\alpha}^{\star}(\Hr)$, which inherently depends on $\Hr$, an outer bound $\bigcup_{\alpha\in[0,1]}\{(\Error^{\rm out}_\alpha,\Rate^{\rm out}_\alpha)\}$ is given by 
\begin{subequations}\label{eqn:outer_bound}
	\begin{align}
		\Rate^{\rm out}_\alpha&=\mathbb{E}\left[\log\det\Bigg(\I+\frac{\HH\tilde{\Rr}_{\X,\alpha}^{\star}(\HH)\HH^\herm}{\sigma_{\rm c}^2}\Bigg)\right];\label{eqn:outer_bound_rate} \\
		\Error^{\rm out}_\alpha &= \mathbb{E}\left\{\Phi\left[\tilde{\Rr}_{\X,\alpha}^{\star}(\HH)\right]\right\}. \label{eqn:outer_bound_error}
	\end{align}
\end{subequations}
\eqref{eqn:outer_bound} is achieved when the input signal $\X$ is simultaneously Gaussian (for the equality in \eqref{eqn:outer_bound_relax_com}
 to hold) and isometry (for the equality in \eqref{eqn:outer_bound_relax_sen} to hold), which is not practically feasible. Hence, \eqref{eqn:outer_bound} is an outer bound.

\paragraph{MMSE-Rate Inner Bounds} Based on the MMSE-Rate outer bound \eqref{eqn:outer_bound}, the following achievable inner bounds may be obtained with the following signaling strategies.

$\divideontimes$ ``$\Ps-\Pc$ \emph{Time-Sharing}" Bound. The two optimal points $\Ps$ and $\Pc$ maybe connected exploiting the time-sharing strategy, which assigns probability $p_{\rm s}\in[0,1]$ to apply the $\Ps-$achieving strategy and probability $p_{\rm c}=1-p_{\rm s}$ to apply the $\Pc-$achieving strategy \cite{FanLiu}. This is a baseline ISAC scheme that splits orthogonal resources between sensing and communication.
	
$\divideontimes$ ``\emph{Communication-Based Inner Bound}''. Motivated by the $\Pc-$achieving strategy, a communication-based inner bound (CIB) $\bigcup_{\alpha\in[0,1]}\{(\Error^{\rm in}_{{\rm c},\alpha},\Rate^{\rm in}_{{\rm c},\alpha})\}$ can be achieved by transmitting the following zero-mean Gaussian signal
\begin{equation}
	\X^{\rm in}_{{\rm c},\alpha}\big|\big(\HH=\Hr\big)=\U_{\!\alpha}(\Hr)\boldsymbol{\mathit\Omega}_\alpha^{1/2}(\Hr)\N,
	\label{eqn:CIB_input}
\end{equation}
for each $\alpha\in[0,1]$, where $\U_{\!\alpha}(\Hr)\in\mathbb{C}^{N\times N}$ and ${\boldsymbol{\mathit\Omega}}_{\alpha}(\Hr)\in\mathbb{R}^{N\times N}$ are from the eigenvalue decomposition of $\tilde{\Rr}_{\X,\alpha}^{\star}(\Hr)$, i.e.,
\begin{equation}
	\tilde{\Rr}_{\X,\alpha}^{\star}(\Hr)=\U_{\!\alpha}(\Hr){\boldsymbol{\mathit\Omega}}_{\alpha}(\Hr)\U_{\!\alpha}^\herm(\Hr).
	\label{eqn:eigen_decomosition_Ralpha}
\end{equation}
The bound is
\begin{subequations}
	\begin{align}
		\Rate^{\rm in}_{{\rm c},\alpha}&= \Rate^{\rm out}_\alpha;\label{eqn:CIB_rate} \\ 
		\Error^{\rm in}_{{\rm c},\alpha} &= \mathbb{E}\Big[\mathbb{E}\left\{{\Phi}\left(\R_{\X}\right)\big|\HH\right\}\Big], \label{eqn:CIB_error}
	\end{align}
\end{subequations}
where the inner expectation is with respect to $\X_{{\rm c},\alpha}^{\rm in}|\HH$, while the outer expectation is with respect to $\HH$.
	
$\divideontimes$ ``\emph{Sensing-Based Inner Bound}''. Motivated by the $\Ps-$achieving strategy, a sensing-based inner bound (SIB) $\bigcup_{\alpha\in[0,1]}\{(\Error^{\rm in}_{{\rm s},\alpha},\Rate^{\rm in}_{{\rm s},\alpha})\}$ can be achieved by transmitting the following isometry signal
\begin{equation}
	{\X^{\rm in}_{{\rm s},\alpha}\big|\big(\HH=\Hr\big)=\sqrt{T}\U_{\!\alpha}(\Hr){\boldsymbol{\mathit\Omega}}_{\alpha}^{1/2}(\Hr)\stiefel}
	\label{eqn:SIB_input}
\end{equation}
for each $\alpha\in[0,1]$, where $\U_{\!\alpha}(\Hr)$ and $\boldsymbol{\mathit\Omega}_\alpha(\Hr)$ are from \eqref{eqn:eigen_decomosition_Ralpha}.
The corresponding high-SNR bound is
\begin{subequations} \label{eqn:SIB}
	\begin{align}
		\Rate^{\rm in}_{{\rm s},\alpha} =&\ \mathbb{E}\left[
		\left(1-\frac{\Ma(\HH)}{2T}\right)\log\det\left(\frac{\HH\tilde{\Rr}_{\X,\alpha}^{\star}(\HH)\HH^\herm}{\sigma_{\rm c}^2}\right)\!+\!{c}_{0,\alpha}(\HH)\right] \notag\\
   &\hspace{0.5cm}+\mathcal{O}(\sigma_{\rm c}^2); \label{eqn:SIB_rate}\\
		\Error^{\rm in}_{{\rm s},\alpha} =&\ \Error^{\rm out}_\alpha, \label{eqn:SIB_error}
	\end{align}
\end{subequations}
where $\Ma(\HH)=\rank(\HH\tilde{\Rr}_{\X,\alpha}^{\star}(\HH)\HH^\herm)$, and $c_{0,\alpha}(\HH)=\frac{\Ma(\HH)}{T}\left[\left(T-\frac{\Ma(\HH)}{2}\right)\log\frac{T}{\e}-\log\Gamma(T)+\log(2\sqrt\pi)\right].$
	
$\divideontimes$ ``\emph{CIB-SIB Time-Sharing}" Bound. A time-sharing inner bound between the CIB and SIB (with respective probability $p_{\rm c}$ to apply the CIB achieving strategy and $p_{\rm s}=1-p_{\rm c}$ to apply the SIB achieving strategy) can be obtained, which is the convex envelope of the CIB and the SIB \cite{FanLiu}.

\begin{remark}
	While \cite{FanLiu} provides foundational inner bounds for the communication–sensing tradeoff under the CRB-Rate framework, our bounds and corresponding achieving strategies for the MMSE-ergodic rate are independently established based on the derived sensing- and communication-optimal waveforms, rather than straightforward extensions of results in \cite{FanLiu}. These bounds reflect the unique requirements on waveform characteristics under the considered MMSE-ergodic rate setup.
\end{remark}

\begin{remark}
	When $T$ is very large, the sample correlation matrix is convergent to the statistical correlation matrix \cite{CRB_BF,CRB_hua}. If $\X^{\rm in}_{{\rm c},\alpha}|\HH$ is transmitted, both equalities in \eqref{eqn:outer_bound_relax} asymptotically hold. Therefore, this outer bound can be asymptotically achieved by \eqref{eqn:CIB_input} for each block when $T$ is large \cite{e26121089}. Reviewing this, the waveform uncertainty tradeoff becomes less important, and the water-filling tradeoff flexibly adjusts the system's ISAC performance when the coherence time is large.
\end{remark}

%% file: numerical_result.tex
\section{Numerical Results}
\label{sec:numerical}
In this section, numerical examples are presented to illustrate the derived theories about the ISAC MMSE-Rate characterization.

\renewcommand{\arraystretch}{1.3}
\begin{table}[t]
	\centering
	\footnotesize
	\caption{Numerical Parameter Default Settings}
	\label{tab:Parameter_Settings}
	\begin{tabular}{lcc}
		\hline
		\text{Parameter} & \text{Symbol} & \text{Value/Range} \\
		\hline
		%Antenna Number of ISAC Tx& $N$ & $6$ \\
		%Antenna Number of Comm. Rx & $\Nc$ & $3$ \\
		%Antenna Number per Sensing Rx & $\Ns'$ & $3$ \\
		%Coherence Time & $T$ & $6$ \\
		B-A Tolerance for the Lagrange Multiplier &$\varepsilon_{\mu}$&$10^{-8}$\\
		B-A Tolerance for ISAC Performance &$\varepsilon_{\!J}$&$10^{-4}$\\
		Comm. Channel Variance & $\sigma_{\mathsf h}^2$ & $1$ \\
		Noise Variance & $\sigma_{\rm s}^2$,$\sigma_{\rm c}^2$ & $\SI{1}{}$ \\
		%Number of Sensing Rx & $R$ & $1\sim7$ \\
		%Transmit Power & $N\Po$ & $\SI{15}{dB}$ \\
		%Transmit Signal-to-Noise Ratio & & $15\sim\SI{20}{dB}$ \\
		\hline
	\end{tabular}
\end{table}

\subsection{Setup}
The default parameter settings for the numerical plot are summarized in Table \ref{tab:Parameter_Settings} unless otherwise stated. The transmit power is defined as $10\log_{10}(N\Po)$, representing the average power of the signal transmitted from the ISAC Tx. The numerical results are obtained in MATLAB, where the numerical optimization problems are solved using the numerical optimizer CVX \cite{CVX}, and the expectation operation is conducted with averaging over $1,000$ independent realizations.

\subsection{Results}

\subsubsection{Single-Input-Single-Output (SISO) ISAC channel}

\begin{figure*}[t]
	\centering
	\definecolor{NUS_blue}{RGB}{0,61,124}
	\definecolor{NUS_orange}{RGB}{239,124,0}
	\boxed{ \textcolor{NUS_blue}{\rule{1cm}{1.5pt}}\ p_{\mathsf{x}|\mathsf{h},\alpha}^\star(x,\hbar) \quad \textcolor{NUS_orange}{\rule{1cm}{1.5pt}}\ p_{\mathsf{y}|\mathsf{h},\alpha}^\star(y,\hbar)}\\
	\vspace{0.1cm}
	\includegraphics[scale=1]{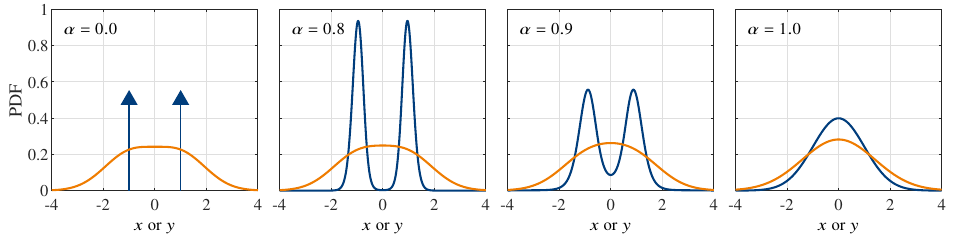}
	\caption{The limit-converging input and output distributions for fast-fading SISO ISAC channels \eqref{eqn:scalar_channel}. The curves are obtained from \eqref{eqn:sensing_opt_waveform} or the B-A algorithm with $\hbar=1$.}
	\label{fig:SISO_distributions}
\end{figure*}

\begin{figure}[t]
	\centering
	\includegraphics[scale=1]{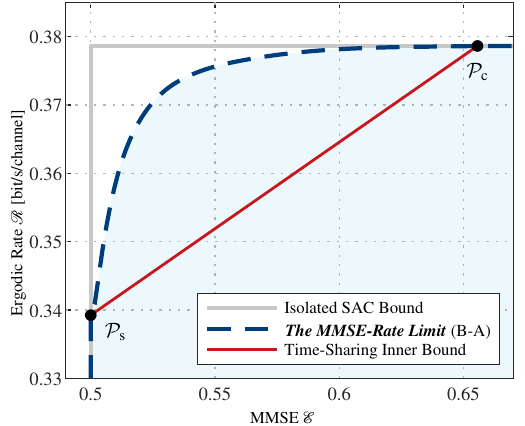}
	\caption{The MMSE-Rate performance limit for fast-fading SISO ISAC channels \eqref{eqn:scalar_channel}.}
	\label{fig:SISO_limit}
\end{figure}

This paper first demonstrates the MMSE-Rate limit for the SISO ISAC channel
\begin{subequations}
	\label{eqn:scalar_channel}
	\begin{align}
		\mathsf{y=hx+z}_{\rm c};\\
		\mathsf{s=gx+z}_{\rm s},
	\end{align}
\end{subequations}
where each symbol is a real-valued random scalar, i.e., $N=\Nc=\Ns=1$, and $T=1$. In this simple example, both the transmit power and the variance of the sensing channel are set to $1$. Given any particular realization of the communication channel $\mathsf{h}=\hbar$, the MMSE-Rate limit-achieving distribution can be numerically obtained based on Algorithm \ref{algorithm:BA}. For example, when $\hbar = 1$, the resulting distribution is illustrated in Fig. \ref{fig:SISO_distributions}. As the preference of functionality of the ISAC signal $\mathsf{x}$ moves from sensing ($\alpha=0$) to communication ($\alpha=1$), the optimal input distribution shifts gradually from a binary distribution with equal probability to a pure Gaussian, and the optimal output distribution transients from a Gaussian convolution with a binary distribution, which has heavier tails, to a pure Gaussian. Intuitively, the ISAC signal must carry more randomness, i.e., become more Gaussian-like, in order to convey more information.

By varying the value of $\hbar$ according to the Rayleigh fading distribution, the entire ISAC limit curve can be obtained, as shown in Fig.~\ref{fig:SISO_limit}. Note that, unlike Fig.~\ref{fig:SISO_distributions} where the channel realization is fixed to $\hbar = 1$, the rate shown here is the ergodic rate.
As expected, the resulting ISAC limit exhibits a concave shape. For the SISO case, the construction of bounds in \cite{FanLiu} can only provide two isolated performance points at $\Ps$ and $\Pc$, as there is no degree of freedom for the intermediate points to select the channel eigendirection and spatial water-filling under SISO. Our bound from the B-A algorithm recovers the full shape of the ISAC limit by computing optimal input distributions at any intermediate tradeoff points.

\begin{figure*}[t]
	\centering
	\includegraphics[scale=1]{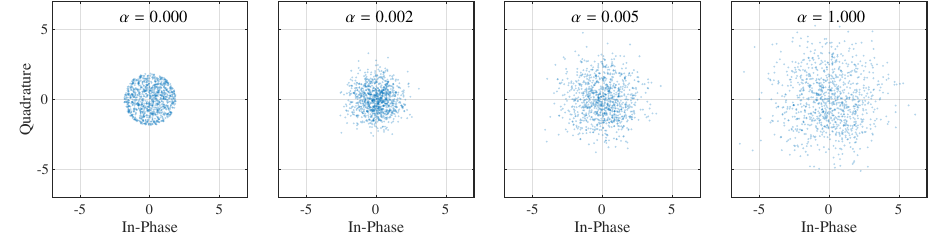}
	\caption{Samples of $[\X]_{1,1}$ generated according to the distribution calculated from \eqref{eqn:sensing_opt_waveform} or using Algorithm \ref{algorithm:BA} with $\Hr=\I_{2}$.}
	\label{fig:MIMO_distribution}
\end{figure*}

\begin{figure}[t]
	\centering
	\includegraphics[scale=1]{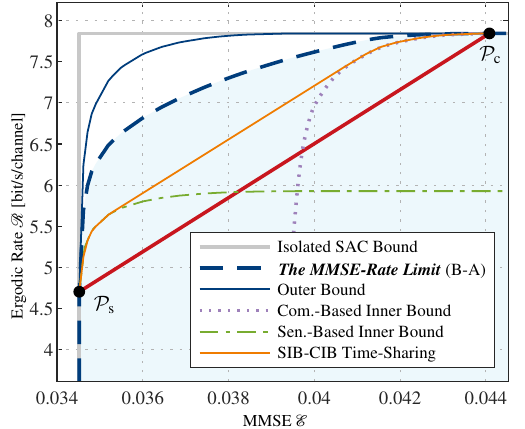}
	\caption{The MMSE-Rate limit and other loose bounds under MIMO channels.}
	\label{fig:MIMO_limit}
\end{figure}

\subsubsection{MIMO ISAC System}
%%% SNR = 15dB;

In this part, the performance of MIMO ISAC channels is presented. Specifically, $N=\Nc=\Ns=2$, and $T=2$. Such a parameter setting is designed to ensure the numerical feasibility of expectation computations over the PDF of the random matrix, which necessitates a reasonable dimensionality, as suggested in \cite{Simulation_parameter_two_channel}. Additionally, the sensing channel $\g$ typically exhibits weak responses due to reflections and long propagation distances. Hence, $\sigma_{\mathsf g}^2=0.03$ is set, $\covr$ and $\covt$ are randomly set (with Hermitian property), and the transmit SNR is set as $\SI{15}{dB}$ in this part.

Given any particular realization of the communication channel $\HH=\Hr$, the MMSE-Rate limit-achieving distribution can be numerically obtained based on Algorithm \ref{algorithm:BA}. For the example of $\Hr = \I_2$, the optimal input behavior is illustrated in Fig.~\ref{fig:MIMO_distribution}, where samples of the $(1,1)$-th entry of $\X$, drawn from the numerically computed densities obtained via Algorithm~\ref{algorithm:BA}, are plotted. It is worth noting that in this setup, the rate significantly dominates the value of MMSE, and thus, noticeable tradeoff behavior only emerges when $\alpha$ is very small. For different values of $\alpha$, the generated samples clearly illustrate the WUT. When $\alpha=0$, the sensing-optimal waveform exhibits high determinism, with all samples confined within a circular disk in the complex plane due to the power budget. As $\alpha$ increases, a smooth transition toward a Gaussian-like distribution is observed, reflecting the waveform structure more favorable for communication.

\begin{figure}[t]
	\centering
	\includegraphics[scale=1]{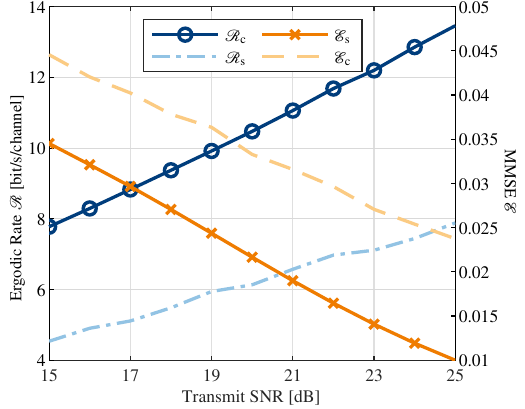}
	\caption{Effects of transmit SNR on sensing- and communication-optimal performance.}
	\label{fig:SNR}
\end{figure}

By varying the values of $\Hr$, the MMSE-Rate performance region can be characterized as in Fig. \ref{fig:MIMO_limit}. In the calculation of sensing-limited rate in the high-SNR regime, the residual term $\mathcal{O}(\sigma_{\rm c}^2)$ in \eqref{eqn:sensing-limited_rate} and \eqref{eqn:SIB} is neglected \cite{FanLiu}. It is observed that both the outer bound and SIB-CIB time-sharing bound fail to accurately characterize the true limit. Note that the CIB-SIB Time-Sharing bound expects better performance compared with $\Ps-\Pc$ Time-Sharing, which allocates orthogonal resources between sensing and communication. Still, it fails to fully leverage the WUT by only considering the combination of sensing- and communication-optimal distributions.

The effects of transmit SNR are demonstrated in Fig. \ref{fig:SNR}. With increased SNR, both performance gets better. Both $\Rc$ and $\Es$ are linear with respect to SNR in $\SI{}{dB}$, as expected. It is also observed that the performance gaps $\Rc-\Rs$ and $\Ec-\Es$ are increasing with high SNR. One possible reason is that at high SNR, the two optimal power allocation schemes $\powers$ and $\powerc$ lose their similarity. Therefore, in the high SNR regime, the WFT will become more prominent and restrict $\Rs$ and $\Ec$.

%% file: conclusion.tex
\section{Conclusion}
\label{sec:conclusion}

This paper has provided a systematic framework for identifying the fundamental performance limits of ISAC systems. Using variational calculus, the optimal distributions on the MMSE-Rate limit were investigated, which are solutions to a high-dimensional complex convolutional equation. A Blahut-Arimoto-type algorithm was proposed to numerically evaluate ISAC performance and limit-achieving distribution. Sensing and communication-optimal points and corresponding achieving strategies were studied, which demonstrate the waveform uncertainty tradeoff and water-filling tradeoff within ISAC. 

% Moreover, a compound signaling strategy for coincided SAC channels was proposed to estimate the channel using the sensing-optimal waveform and to convey information using the communication-optimal waveform over the estimated channel. It demonstrates the potential for significant performance improvements by integrating sensing and communication.

%Designing the computation effective algorithms for numerically compute the ISAC limit for large MIMO systems will be subject to future research. 
Overall, this study presents an approach to identify the joint estimation- and information-theoretic performance limits for ISAC systems. The theoretical findings in this work highlight both gains and tradeoffs between sensing and communication within an integrated waveform. It should be noted that two tasks remain: further characterizing the properties of Pareto-optimal distributions and analytically exploring the Pareto boundary. Hopefully, this work will provide valuable insights into practical ISAC system performance characterization and design for future wireless networks. 

%% file: Appendix/Appendix1.tex
\appendices
\numberwithin{equation}{section}

\section{Proof of Theorem \ref{theorem:limmit_distribution}}
\label{appendix:limit}
With communication channel realization $\HH=\Hr$, the input-output relationship of the MIMO communication channel reduces to $\Y=\Hr\X+\Zc$. With a given $\alpha\in[0,1]$, the conditional mutual information is
\begin{align*}
	&\ \mathcal{I}(\X;\Y|\HH=\Hr)\\
	=&\ h(\Y|\HH=\Hr) - h(\Y|\X,\HH=\Hr) \\
	=&\ -\int \PY\log\PY\ \diff\Yr - h(\Zc).
\end{align*}
Note that $h(\Zc)$ is independent of $p_{\X|\HH}$ and $p_{\Y|\HH}$. Therefore, \eqref{eqn:ISAC_limit_SAA} reduces to
\begin{subequations} \label{eqn:functional_opt_intitial}
	\begin{align}
		\sup_{\substack{p_{\X|\HH}\\p_{\Y|\HH}}}&\quad -\frac\alpha T\!\! \int\!\! \PY\log \PY\ \diff\Yr \notag \\
        &\hspace{1.3cm} -(1-\alpha)\!\!\int\!\! \PX \Phi(\Rr_{\Xr})
			\ \diff\Xr\\
		\st& \quad \!\!\int\!\! \PX\ \diff\Xr = 1, \label{eqn:variation_unity}\\
		& \quad \!\!\int\!\! \PX {\tr}\left(\Xr\Xr^\herm\right)\ \diff\Xr \leq TN\Po,  \label{eqn:variation_power}\\
		& \quad  \!\!\int\!\! \PY\ \diff\Yr = 1, \\
		& \quad  \PY = \!\!\int\!\! \PX p_{\Zc}(\Yr-\Hr\Xr)\ \diff\Xr,
	\end{align}
\end{subequations}
where $p_{\Zc}(\cdot)$ is the distribution of the communication noise $\Zc$, which is assumed to be Gaussian. Expanding $\PX=\int \PX p_{\Zc}(\Yr-\Hr\Xr)\diff\Yr$, \eqref{eqn:functional_opt_intitial} reduces to
\begin{subequations}
	\begin{align}
		\inf_{\substack{p_{\X|\HH}\\p_{\Y|\HH}}}& \quad \!\!\iint\!\! \PX p_{\Zc}(\Yr-\Hr\Xr)\Big[\frac\alpha T\log\PY \notag \\
        & \quad\quad\quad + (1-\alpha)\Phi(\Rr_{\Xr})
		\Big]\ \diff\Xr\diff\Yr\\
		\st& \iint\!\! \PX p_{\Zc}(\Yr-\Hr\Xr)\ \diff\Xr\diff\Yr = 1, \label{eqn:functional_opt_pdf}\\
		& \iint\!\! \PX p_{\Zc}(\Yr-\Hr\Xr){\tr}\big(\Xr\Xr^\herm\big)\ \diff\Xr\diff\Yr \leq TN\Po, \label{eqn:functional_opt_power} \\
		%& \quad  \int\PY\diff\Yr = 1, \label{eqn:functional_opt_y_pdf}\\
		& \PY = \!\!\int\!\! \PX p_{\Zc}(\Yr-\Hr\Xr)\ \diff\Xr, \label{eqn:functional_conv}
	\end{align}
	\label{eqn:functional_opt}
\end{subequations}
which can be efficiently solved with the method of variational calculus.

Using \cite[Corollary 1]{variation}, the functional optimization problem \eqref{eqn:functional_opt} is equivalent to
\begin{equation*}
    \begin{aligned}
        \inf_{\substack{p_{\X|\HH}\\p_{\Y|\HH}}} \quad  & U[p_{\X|\HH},p_{\Y|\HH}] \\
        =& \iint\!\!\Big[K_\alpha(\Xr,\Yr,p_{\X|\HH},p_{\Y|\HH})\ \diff\Xr+\tilde{K}_\alpha(\Yr,p_{\Y|\HH})\Big]\ \diff\Yr,
    \end{aligned}
\end{equation*}
where
\begin{align}
    & K_\alpha(\Xr,\Yr,p_{\X|\HH},p_{\Y|\HH}) \notag \\
    & \ \ = \PX p_{\Zc}(\Yr -\Hr\Xr)\Big[\frac\alpha T \log\PY \notag \\ 
    & \ \ \quad +(1-\alpha)\Phi(\Rr_{\Xr}) +\tilde\mu_{1,\alpha}+\tilde\mu_{2,\alpha}\tr(\Xr\Xr^\herm)-\tilde\mu_{3,\alpha}\Big]; \label{eqn:functional_K} \\ 
    & \tilde{K}_\alpha(\Yr,p_{\Y|\HH})=\tilde\mu_{3,\alpha}\PY.
	\label{eqn:functional_K'}
\end{align}
In \eqref{eqn:functional_K} and \eqref{eqn:functional_K'}, $\tilde\mu_{1,\alpha}$, $\tilde\mu_{2,\alpha}$ and $\tilde\mu_{3,\alpha}$ are respectively the Lagrange multipliers associated with the constraints \eqref{eqn:functional_opt_pdf}, \eqref{eqn:functional_opt_power} and \eqref{eqn:functional_conv}. Based on \cite[Corollary 2]{variation}, confirming the first-order variation conditions, the optimal solutions must satisfy the Euler-Lagrange equations given by 
\begin{equation*} 
    \begin{aligned}
        & \left\{
            \begin{aligned}
                &p^\star_{\X|\HH,\alpha}\\
                &p^\star_{\Y|\HH,\alpha}
            \end{aligned}
        \right\} = \\
        &\ \underset{\substack{p_{\X|\HH}\\p_{\Y|\HH}}}{\arg}
          \left\{
            \begin{aligned}
                & \int \frac{\delta  K_\alpha(\Xr,\Yr,p_{\X|\HH},p_{\Y|\HH})}{\delta p_{\X|\HH}}\ \diff \Yr=0 \\
                & \int \frac{\delta  K_\alpha(\Xr,\Yr,p_{\X|\HH},p_{\Y|\HH})}{\delta p_{\Y|\HH}}\ \diff \Xr +\frac{\delta \tilde{K}_\alpha(\Yr,p_{\Y|\HH})}{\delta p_{\Y|\HH}} =0 
            \end{aligned}
          \right\}.
    \end{aligned}
\end{equation*}
Explicitly solving them yields $\tilde\mu_{3,\alpha}=-\alpha/T$, and \eqref{eqn:opt_out_distribution}, where $\mu_{1,\alpha}=T[\tilde{\mu}_{3,\alpha}-\tilde{\mu}_{1,\alpha}]$, and $\mu_{2,\alpha}=-T\tilde{\mu}_{2,\alpha}$ are the transformed Lagrange multipliers for simplicity. Confirming the second-order variational condition, it can be verified that the solutions $p^\star_{\X|\HH,\alpha}$ and $p^\star_{\Y|\HH,\alpha}$ indeed achieve the maximum/supremum of the functional optimization problem. \qed

%%%%%%%%%%%%%%%%%%%%
%%% un-achievability
%%%%%%%%%%%%%%%%%%%%

\section{Proof of Theorem \ref{theorem:unachivability}}
\label{appendix:unachievability}
\begin{proof}
		$\divideontimes$ When $\alpha=0$, \eqref{eqn:opt_out_distribution} reduces to
		\begin{equation*}
			-T\Phi(\Rr_{\Xr})+\mu_{1,0}+ \mu_{2,0}\tr(\Xr\Xr^\herm)=0,
		\end{equation*}
		which is an equation with respect to $\Rr_{\Xr}=\frac1T \Xr\Xr^\herm$ (or $\X\X^\herm$). In order for this equation to consistently hold despite $\X$ being random, the matrix $\X\X^\herm$ must be restricted to the set of solution(s) to the above equation. Hence, the support is not the whole space $\mathbb{C}^{N\times T}$, and \eqref{eqn:opt_out_distribution} can not be satisfied by any PDF in $\mathcal{C}$.

		For $\alpha\neq 0$, multi-dimensional Hermite polynomial decomposition is adopted to address \eqref{eqn:opt_out_distribution}.
		
		\begin{definition}[Hermite polynomials \cite{Hermite_book}]
			A multi-dimensional Hermite polynomial of degree $n$ is
			\begin{equation*}
				\mathcal{H}^{(n)}_{\lambda_1\cdots\lambda_n}(\yr):=(-1)^n\e^{\frac{\|\yr\|^2}{2}}\frac{\partial}{\partial y_{\lambda_1}}\cdots\frac{\partial}{\partial y_{\lambda_n}} \e^{-\frac{\|\yr\|^2}{2}},
			\end{equation*}
			with $\lambda_i\in\mathbb{Z}^+$ and $\lambda_i\leq {\rm dim}(\yr)$.
		\end{definition}

		In sequel, Hermite polynomials up to order $n=2$ will be used for proof, which are readily given by \cite{Hermite_book}
		\begin{equation*}
			\mathcal{H}^{(0)}(\yr)=1;\ \mathcal{H}^{(1)}_{\lambda_1}(\yr)= y_{\lambda_1};\ \mathcal{H}^{(2)}_{\lambda_1\lambda_2}(\yr)=y_{\lambda_1}y_{\lambda_2}-\delta_{\lambda_1\lambda_2}.
		\end{equation*}

		The channel input-output relation $\Yr=\Hr\Xr+\boldsymbol{Z}_{\rm c}$ can be rearranged into real-valued form of $\yr=\tilde{\Hr}\xr+\boldsymbol{z}_{\rm c}$, where
        \begin{align*}
            \yr&:=[{\rm vec}(\mathfrak{Re}(\Yr))^\trans,{\rm vec}(\mathfrak{Im}(\Yr))^\trans]^\trans& &\in\mathbb{R}^{\Nc T\times 1}; \\
            \xr&:=[{\rm vec}(\mathfrak{Re}(\Xr))^\trans,{\rm vec}(\mathfrak{Im}(\Xr))^\trans]^\trans& &\in \mathbb{R}^{NT\times 1}; \\
            \boldsymbol{z}_{\rm c}&:=[{\rm vec}(\mathfrak{Re}(\boldsymbol{Z}_{\rm c}))^\trans,{\rm vec}(\mathfrak{Im}(\boldsymbol{Z}_{\rm c}))^\trans]^\trans& &\in \mathbb{R}^{\Nc T\times 1}; \\
            \tilde{\Hr}&:=
                \begin{bmatrix}
                \mathfrak{Re}(\I_T \otimes \Hr) & -\mathfrak{Im}(\I_T \otimes \Hr) \\
                \mathfrak{Im}(\I_T \otimes \Hr) & \mathfrak{Re}(\I_T \otimes \Hr)
                \end{bmatrix}& &\in\mathbb{R}^{\Nc T\times NT}.
        \end{align*}
Note that this rearrangement is a one-to-one mapping. Hence, the distribution of $\X$, $\Y$, and $\Zc$ can be fully described by functions with $\xr$, $\yr$, and $\boldsymbol{z}_{\rm c}$ as variables, respectively. Also note that no matter what the distribution of $\X$ is, the distribution of $\Y$, being the convolution of $\X$ with a Gaussian distribution, is continuous, smooth, and differentiable. Consequently, the function $\log \PYopt$ can be decomposed as
	\begin{equation}
		\log \PYopt=\sum_{n=0}^\infty\sum_{\pmb{\lambda}_n} a_{\pmb{\lambda}_n}^{(n)}\mathcal{H}_{\pmb{\lambda}_n}^{(n)}(\yr),
		\label{eqn:Hermite}
	\end{equation}
	where $\pmb{\lambda}_n$ is any proper permutation of $\lambda_1\lambda_2\cdots\lambda_n$, and $a_{\pmb{\lambda}_n}^{(n)}$ is the associated Hermite coefficient. According to the property of the Hermite polynomial (see details in \cite{Hermite_book,Hermite_CL}),
		\begin{equation*}
			\begin{aligned}
				\int p_{\Zc}(\Yr-\Hr\Xr)&\log \PYopt\ \diff\Yr\\
				&=\sum_{n=0}^\infty  \sum_{\pmb{\lambda}_n}a_{\pmb{\lambda}_n}^{(n)}\prod_{i=1}^{2\Nc T} (\tilde{\hr}_i^{\trans}\xr)^{\sum_j \delta([\pmb{\lambda}_n]_j-i)},
			\end{aligned}
		\end{equation*}
		where $\tilde{\boldsymbol{h}}_i^\trans=\boldsymbol{e}_i^\trans\tilde{\Hr}$ stands for the $i$-th row of $\tilde{\Hr}$, with $\boldsymbol{e}_i$ denoting the $i$-th standard orthonormal Euclidean basis.

$\divideontimes$ When $\alpha=1$, \eqref{eqn:opt_out_distribution} reduces to
$$\begin{aligned}
		& \mu_{1,1}+ \mu_{2,1}\xr^\trans\xr\\
        =&\!\!\int\!\! p_{\Zc}(\Yr-\!\Hr\Xr)\log p_{\Y|\HH,1}^\star(\Yr|\Hr)\ \diff\Yr\\
				=&\ a^{(0)}_{\boldsymbol{0}}+a^{(1)}_{[1,0,\cdots,0]}(\tilde{\hr}_1^\trans\xr)+\cdots\\
				& +a^{(1)}_{[0,0,\cdots,1]}(\tilde{\hr}_{2\Nc T}^\trans\xr) + a^{(2)}_{[2,0,\cdots,0]}(\tilde{\hr}_{1}^\trans\xr)^2+\cdots\\
				& +a^{(2)}_{[0,0,\cdots,2]}(\tilde{\hr}_{2\Nc T}^\trans\xr)^2+a^{(2)}_{[1,1,\cdots,0]}(\tilde{\hr}_{1}^\trans\xr\cdot \tilde{\hr}_{2}^\trans\xr)+ \cdots \\
				& +a^{(2)}_{[0,\cdots,1,1]}(\tilde{\hr}_{2\Nc T-1}^\trans\xr\cdot \tilde{\hr}_{2\Nc T}^\trans\xr) + \mathcal{O}(x^3).
\end{aligned}$$
	This equation holds for every $\xr\in\mathbb{R}^{2NT\times 1}$.
	Comparing both sides and matching the coefficient for each momentum of $\xr$, it can be seen that only the zero-th and second order terms with respect to $\xr$ should be present, i.e., $a^{(n)}_{\pmb{\lambda}_n}=0$ for $n\in\mathbb{Z}_+\backslash\{2\}$. Consequently,
	$$\log p_{\Y|\HH,1}^\star(\Yr|\Hr)=a_{\boldsymbol{0}}^{(0)}\mathcal{H}_{\pmb{\lambda}_0}^{(0)}(\yr)+\sum_{\pmb{\lambda}_2}a_{\pmb{\lambda}_2}^{(2)}\mathcal{H}_{\pmb{\lambda}_2}^{(2)}(\yr).$$
	This function is entirely determined by the second momentum of $\yr$; it does not require any first-order or higher-order statistics or distributional details. Consequently, $p_{\Y|\HH,1}^\star$ is a multivariate Gaussian. Since $\Zc$ is a Gaussian, $p_{\X|\HH,1}^\star$ is a Gaussian as well.

$\divideontimes$ When $\alpha\in(0,1)$, it is to be shown that there does not exist a $\PYopt$ such that \eqref{eqn:opt_out_distribution} holds for each $\Xr\in\mathbb{C}^{N\times T}$ (since $\mathcal{C}$ is defined on PDFs with full support). This is proven by contradiction: suppose $\PYopt$ exists such that \eqref{eqn:opt_out_distribution} holds for each $\Xr$, then \eqref{eqn:opt_out_distribution} must hold for any rank-1 matrix $\Xr=\uu\boldsymbol{e}_1^\trans$. From the Sherman-Morrison identity,
\begin{equation*}
	\begin{aligned}
		&\ (\cov^{-1}+\sigma_{\rm s}^{-2}\I_{\Ns}\otimes\Xr\Xr^\herm)^{-1} \\
		=&\ (\cov^{-1}+\sigma_{\rm s}^{-2}\I_{\Ns}\otimes\boldsymbol{uu}^\herm)^{-1}\\
		=&\ \cov-\cov(\I_{\Ns}\otimes \uu)(\sigma_{\rm s}^2\I_{\Ns}+\sigma_{\mathsf g}^2\uu^\herm \covt \uu\covr^\trans)^{-1}(\I_{\Ns}\otimes \uu^\herm)\cov.
	\end{aligned}
\end{equation*}
Hence,
\begin{equation*}
	\begin{aligned}
		&\ \Phi(\RXr) = \tr(\cov^{-1}+\sigma_{\rm s}^{-2}\I_{\Ns}\otimes\Xr\Xr^\herm)^{-1}\\
		&= \tr(\cov) \\
		& \quad\ \ -\tr[\cov(\I_{\Ns}\otimes \uu)(\sigma_{\rm s}^2\I_{\Ns}+\sigma_{\mathsf g}^2\uu^\herm \covt \uu\covr^\trans)^{-1}(\I_{\Ns}\otimes \uu^\herm)\cov]\\
		&= \tr(\cov) - \sigma_{\mathsf g}^{4}\uu^\herm \covt^2 \uu \tr[(\sigma_{\rm s}^2\I_{\Ns}+\sigma_{\mathsf g}^2\uu^\herm\covt\uu\covr^\trans)^{-1}(\covr^{\trans})^2]\\
		&= \tr(\cov)-\sigma_{\mathsf g}^4 \sum_{i=1}^{\Ns} \frac{\eigr{i}^2\uu^\herm\covt^2\uu}{\sigma_{\rm s}^2+\sigma_{\mathsf g}^2\eigr{i}\uu^\herm\covt\uu}.
	\end{aligned}
\end{equation*}

To solve \eqref{eqn:opt_out_distribution}, the above equation must be written as a function of $\xr$. Note that
	$$
	\begin{aligned}
		\uu^\herm \covt\uu &= [\mathfrak{Re}(\uu)+\jj\mathfrak{Im}(\uu)]^\herm \covt [\mathfrak{Re}(\uu)+\jj\mathfrak{Im}(\uu)]\\
		&= \mathfrak{Re}(\uu)^\trans\covt\mathfrak{Re}(\uu)+\mathfrak{Im}(\uu)^\trans\covt\mathfrak{Im}(\uu)\\
		&\quad + \jj [\mathfrak{Re}(\uu)^\trans\covt\mathfrak{Im}(\uu)-\mathfrak{Im}(\uu)^\trans\covt\mathfrak{Re}(\uu)]\\
		&= \mathfrak{Re}(\uu)^\trans\covt\mathfrak{Re}(\uu)+\mathfrak{Im}(\uu)^\trans\covt\mathfrak{Im}(\uu)\\
		&= \xr^\trans \tilde{\boldsymbol{\mathit\Sigma}}_{\rm t}\xr,
	\end{aligned}
	$$
where $\tilde{\boldsymbol{\mathit\Sigma}}_{\rm t} :=\mathrm{diag}(\covt,\boldsymbol{0}_{N(T-1)\times N(T-1)},\covt,\boldsymbol{0}_{N(T-1)\times N(T-1)})$. Similarly, $\uu^\herm\covt^2\uu=\xr^\trans \tilde{\boldsymbol{\mathit\Sigma}}_{\rm t}^2 \xr$. Hence,
	$$\Phi(\Rr_{\Xr}) = \tr(\cov)-\sigma_{\mathsf g}^4 \sum_{i=1}^{\Ns} \frac{\eigr{i}^2\xr^\trans \tilde{\boldsymbol{\mathit\Sigma}}_{\rm t}^2\xr}{\sigma_{\rm s}^2+\sigma_{\mathsf g}^2\eigr{i}\xr^\trans \tilde{\boldsymbol{\mathit\Sigma}}_{\rm t}\xr}.$$
Similar to the $\alpha=1$ case, in order to solve for $\PYopt$, the Hermite coefficient of $\log \PYopt$ in \eqref{eqn:Hermite} can be identified by matching the coefficients related to each momentum of $\xr$ on both sides of \eqref{eqn:opt_out_distribution}.

Note that the power expansion of $\Phi(\Rr_{\Xr})$ with respect to $\xr$ is convergent only when ($\eigr{1}$ is the largest eigenvalue of $\covr$)
$$\left|\frac{\sigma_{\mathsf g}^2}{\sigma_{\rm s}^{2}}\eigr{1}\xr^\trans \tilde{\boldsymbol{\mathit\Sigma}}_{\rm t} \xr \right|<1.$$
Consequently, when $\Big|\frac{\sigma_{\mathsf g}^2}{\sigma_{\rm s}^{2}}\eigr{1}\xr^\trans \tilde{\boldsymbol{\mathit\Sigma}}_{\rm t} \xr\Big|\geq1$, the Hermite coefficients can not be determined as the right-hand side of \eqref{eqn:opt_out_distribution} will be divergent, and \eqref{eqn:opt_out_distribution} can not be satisfied.

Since $\PYopt$ should always satisfy \eqref{eqn:opt_out_distribution} regardless of the choice of $\Xr=\uu\boldsymbol{e}_1^\trans$, no optimal solution $\PYopt$ exists when $\Xr$ is a rank-1 matrix. This contradicts the assumption that ``$\PYopt$ exists such that \eqref{eqn:opt_out_distribution} holds for every $\Xr$''. By contradiction, $\PYopt$ does not exist such that \eqref{eqn:opt_out_distribution} holds for each $\Xr$ with $\alpha\in(0,1)$. This implies that $\PXopt$ does not exist in $\mathcal{C}$ for $\alpha\in(0,1)$. 
\end{proof}

\begin{remark}
	This proof is constructed using a rank-one matrix as a counterexample. Intuitively, after applying the Hermite expansion, the left-hand side of equation \eqref{eqn:opt_out_distribution} becomes a polynomial function of $\Xr$. In contrast, the $\Phi(\Rr_{\Xr})$ term on the right-hand side, involving a matrix inverse, cannot be globally represented as a power expansion of $\Xr$. Due to their different functional forms, it is unlikely that the two sides can coincide for all $\Xr \in \mathbb{C}^{N \times T}$.
\end{remark}

%%%%%%%%%%%%%%%%%%%%
%%% BA-algorithm
%%%%%%%%%%%%%%%%%%%%

\section{Blahut-Arimoto Type Algorithm\\ for Evaluating the MMSE-Rate Limit}
\label{appendix:BA}
\subsection{Derivation}
	With communication channel realization $\HH=\Hr$, the input-output relationship of the MIMO communication channel reduces to $\Y=\Hr\X+\Zc$. The conditional mutual information is rewritten as
	\begin{equation}
		\begin{aligned}
				& \mathcal{I}(\X;\Y|\HH=\Hr) \\
			=& \int\!\! p_{\X,\Y|\HH}(\Xr,\Yr|\Hr)\log\frac{p_{\X,\Y|\HH}(\Xr,\Yr|\Hr)}{p_{\X|\HH}(\Xr|\Hr)p_{\Y|\HH}(\Yr|\Hr)}\ \diff\Xr\diff\Yr \\
			=& \int\!\! \PX p_{\Y|\X,\HH}(\Yr|\Xr,\Hr)\log\frac{\phi(\Xr|\Yr,\Hr)}{\PX}\ \diff\Xr\diff\Yr, 
		\end{aligned}
	\end{equation}
	where $\phi(\Xr|\Yr,\Hr)$ is the conditional probability (transition probability from the channel output to input).

	With a given $\alpha\in(0,1]$, \eqref{eqn:ISAC_limit_SAA} reduces to
	\begin{subequations}
		\begin{align}
			\sup_{p_{\X}}&\quad \frac{\alpha}{T}\int \PX p_{\Y|\X,\HH}(\Yr|\Xr,\Hr)\log\frac{\phi(\Xr|\Yr,\Hr)}{\PX}\ \diff\Xr\diff\Yr \notag \\
			&\hspace{1cm} -(1-\alpha)\int \PX \Phi(\Rr_{\Xr})\ \diff\Xr \label{eqn:single_max_obj}\\
			\st & \quad {\text{\eqref{eqn:variation_unity} and \eqref{eqn:variation_power}}}  \label{eqn:single_max}
		\end{align}
		\label{eqn:single_maximization}
	\end{subequations}
	Applying the alternating double maximization principle, the objective functional \eqref{eqn:single_max_obj} reduces to
	\begin{equation} 
		\sup_{p_{\X|\HH}}\ \sup_{\phi} \quad \breve{J}(p_{\X|\HH},\phi)=\int J(p_{\X|\HH},\phi,\Xr)\ \diff\Xr,
        \label{eqn:double_maximization}
	\end{equation}
    with
    \begin{equation} \label{eqn:J_intergrate_X}
        \begin{aligned}
			& J(p_{\X|\HH},\phi,\Xr)\\
			=& \int\frac{\alpha}{T} \PX p_{\Y|\X,\HH}(\Yr|\Xr,\Hr)\log\frac{\phi(\Xr|\Yr,\Hr)}{\PX}\ \diff\Yr \\
		      &\ -(1-\alpha) \PX\Phi(\Rr_{\Xr}) \\
		      &\ +\nu\phi(\Xr|\Yr,\Hr)+\tilde\mu_1 \PX \\
            &\ +\tilde\mu_2\PX \tr(\Xr\Xr^\herm),
	       \end{aligned}
    \end{equation}
	and $\nu$, $\tilde\mu_1$ and $\tilde{\mu}_2$ are the associated Lagrange multipliers. For a fixed $p_{\X|\HH}$, the optimal transition probability $\phi^\star$ is
	\begin{align}
		\phi^\star&=\arg_{\phi}\left\{\frac{\delta}{\delta \phi}J(p_{\X|\HH},\phi,\Xr)=0\right\} \notag \\
		&=-\frac{\alpha \PX p_{\Y|\X,\HH}(\Yr|\Xr,\Hr)}{T\nu}\\
		&=\frac{\PX p_{\Y|\X,\HH}(\Yr|\Xr,\Hr)}{\int \PX p_{\Y|\X,\HH}(\Yr|\Xr,\Hr)\ \diff\Xr}. \notag
	\end{align}

\begin{figure*}[t]
	\begin{equation} \label{eqn:opt_input_distribution} 
		p_{\X|\HH}^\star=\frac{\exp\left[\int p_{\Y|\X,\HH}(\Yr|\Xr,\Hr)\log\phi^\star\ \diff \Yr\right]}{\exp\left[1-\mu_1-\mu_2\tr(\Xr\Xr^\herm)+(\frac1\alpha-1)T\Phi(\Rr_{\Xr})\right]}
		=\ \frac{p_{\X|\HH}^\star\exp\left[\int p_{\Y|\X,\HH}(\Yr|\Xr,\Hr)\log\frac{p_{\Y|\X,\HH}(\Yr|\Xr,\Hr)}{\int p_{\X|\HH}^\star p_{\Y|\X,\HH}(\Yr|\Xr,\Hr)\ \diff\Xr}\ \diff \Yr\right]}{\exp(1-\mu_1)\exp\left[-\mu_2\tr(\Xr\Xr^\herm)+(\frac1\alpha-1)T\Phi(\Rr_{\Xr})\right]}.
	\end{equation}
	\rule{\linewidth}{0.1pt}
	\begin{equation} \label{eqn:BA_Lagrange} 
		\begin{aligned}
			\int \frac{p_{\X|\HH}^\star\exp\left[\int p_{\Y|\X,\HH}(\Yr|\Xr,\Hr)\log\frac{p_{\Y|\X,\HH}(\Yr|\Xr,\Hr)}{\int p_{\X|\HH}^\star p_{\Y|\X,\HH}(\Yr|\Xr,\Hr)\ \diff\Xr}\ \diff \Yr\right]}{\exp(1-\mu_1)\exp\left[-\mu_2\tr(\Xr\Xr^\herm)+(\frac1\alpha-1)T\Phi(\Rr_{\Xr})\right]}\ \diff\Xr&=1;\\
			\int\tr(\Xr\Xr^\herm)\frac{p_{\X|\HH}^\star\exp\left[\int p_{\Y|\X,\HH}(\Yr|\Xr,\Hr)\log\frac{p_{\Y|\X,\HH}(\Yr|\Xr,\Hr)}{\int p_{\X|\HH}^\star p_{\Y|\X,\HH}(\Yr|\Xr,\Hr)\ \diff\Xr}\ \diff \Yr\right]}{\exp(1-\mu_1)\exp\left[-\mu_2\tr(\Xr\Xr^\herm)+(\frac1\alpha-1)T\Phi(\Rr_{\Xr})\right]}\ \diff\Xr&=NT\Po.
		\end{aligned}
	\end{equation}
	\rule{\linewidth}{0.1pt}
	\begin{equation} \label{eqn:def_f}
		f(\mu_2):=\int \left(1-\frac{\tr(\Xr\Xr^\herm)}{NT\Po}\right)\frac{p_{\X|\HH}^\star\exp\left[\int p_{\Y|\X,\HH}(\Yr|\Xr,\Hr)\log\frac{p_{\Y|\X,\HH}(\Yr|\Xr,\Hr)}{\int p_{\X|\HH}^\star p_{\Y|\X,\HH}(\Yr|\Xr,\Hr)\ \diff\Xr}\ \diff \Yr\right]}{\exp\left[(\frac1\alpha-1)T\Phi(\Rr_{\Xr})\right]}\exp\left[\mu_2\tr(\Xr\Xr^\herm)\right]\ \diff\Xr.
	\end{equation}
	\rule{\linewidth}{0.1pt}
\end{figure*}

	Given $\phi^\star$, the optimal input probability $p_{\X|\HH}^\star$ is
	\begin{equation*}
		p_{\X|\HH}^\star =\arg_{p_{\X|\HH}}\left\{ \frac{\delta}{\delta p_{\X|\HH}} {J}(p_{\X|\HH},\phi^\star,\Xr)=0 \right\},
	\end{equation*}
	which yields \eqref{eqn:opt_input_distribution}, wherein $\mu_1=\frac T\alpha\tilde{\mu}_1$, and $\mu_2=\frac T\alpha\tilde{\mu}_2$ for simplicity. To solve for the associated Lagrange multipliers, plugging \eqref{eqn:opt_input_distribution} into the constraints \eqref{eqn:single_max} yields \eqref{eqn:BA_Lagrange}. Eliminating the common scaling factor $\exp(1-\mu_1)$ from above two equations yields $f(\mu_2)=0$ with $f(\cdot)$ defined as \eqref{eqn:def_f}.
	It is observed that $f(\mu_2)=0$ is a non-linear equation with respect to $\mu_2$. Numerically, $\mu_2$ can be solved using the Newton-Raphson method, yielding the iterative solution
	\begin{equation*}
		\mu_2^{(j)}=\mu_2^{(j-1)}-\frac{f(\mu_2^{(j-1)})}{f'(\mu_2^{(j-1)})}.
	\end{equation*}
	Once $\mu_2$ is obtained, $p_{\X|\HH}^\star$ can be iteratively calculated from \eqref{eqn:opt_input_distribution}. \qed

	\subsection{Proof of Convergence}
	First, it is shown that the functional $\breve{J}(p_{\X|\HH},\phi)$ in \eqref{eqn:double_maximization} is concave with respect to the ordered pair $(p_{\X|\HH},\phi)$.
	Denote two ordered functional pairs as $(p_{\X|\HH}^{(1)},\phi^{(1)})$, $(p_{\X|\HH}^{(2)},\phi^{(2)})$, and $\kappa\in[0,1]$, $\bar{\kappa}:=1-\kappa$. According to the log sum inequality \cite[Theorem 2.7.1]{cover1999elements}
	\begin{equation*}
		\begin{aligned}
			&\ \frac{\alpha}{T}(\kappa p_{\X|\HH}^{(1)}+\bar\kappa p_{\X|\HH}^{(2)})\log\frac{\kappa p_{\X|\HH}^{(1)}+\bar\kappa p_{\X|\HH}^{(2)}}{\kappa\phi^{(1)}+\bar\kappa\phi^{(2)}}
			\\
			& \hspace{1cm} +(1-\alpha)(\kappa p_{\X|\HH}^{(1)}+\bar\kappa p_{\X|\HH}^{(2)})\Phi(\Rr_{\Xr})\\
			\leq&\ \kappa \frac{\alpha}{T} p_{\X|\HH}^{(1)}\log\frac{p_{\X|\HH}^{(1)}}{\phi^{(1)}}+\bar\kappa \frac{\alpha}{T} p_{\X|\HH}^{(2)}\log\frac{p_{\X|\HH}^{(2)}}{\phi^{(2)}} \\
			&\hspace{1cm} +\kappa(1-\alpha)p_{\X|\HH}^{(1)}\Phi(\Rr_{\Xr})+\bar\kappa(1-\alpha)p_{\X|\HH}^{(2)}\Phi(\Rr_{\Xr}).
		\end{aligned}
	\end{equation*}
	Multiplying both sides with $-p_{\Y|\X,\HH}(\Yr|\Xr,\Hr)$ and integrating with respect to $\Yr$ yield
	\begin{equation*}
		\begin{aligned}
			&J(\kappa p_{\X|\HH}^{(1)}+\bar\kappa p_{\X|\HH}^{(2)},\kappa\phi^{(1)}+\bar\kappa\phi^{(2)},\Xr)-\nu(\kappa\phi^{(1)}+\bar\kappa\phi^{(2)}) \\
			\geq\ & \kappa[ J(p_{\X|\HH}^{(1)},\phi^{(1)},\Xr)-\nu\phi^{(1)}]+\bar\kappa[ J(p_{\X|\HH}^{(2)},\phi^{(2)},\Xr)-\nu\phi^{(2)}],
		\end{aligned}
	\end{equation*}
	where $J(\cdot)$ is given in \eqref{eqn:J_intergrate_X}. Integrating both sides with $\Xr$ and simplifying yield
	\begin{equation*}
		\begin{aligned}
			&\breve{J}(\kappa p_{\X|\HH}^{(1)}+\bar\kappa p_{\X|\HH}^{(2)},\kappa \phi^{(1)}+\bar\kappa \phi^{(2)})\\
			\geq\ & \kappa\breve{J}(p_{\X|\HH}^{(1)},\phi^{(1)})+\bar\kappa\breve{J}(p_{\X|\HH}^{(2)},\phi^{(2)}).
		\end{aligned}
	\end{equation*}
	Therefore, $\breve{J}$ is concave with respect to the ordered pair $(p_{\X|\HH},\phi)$. According to \cite[Theorem 9.5]{yeung2008information}, $$\breve{J}(p_{\X|\HH}^{(\infty)},\phi^{(\infty)})\to\sup_{p_{\X|\HH},\phi}\breve{J}(p_{\X|\HH},\phi),$$ showing the procedures provided in Algorithm \ref{algorithm:BA} will be convergent to the MMSE-Rate limit.
\qed

%% file: Appendix/Appendix2.tex
\section{Proof of Theorem \ref{theorem:sensing_opt}}
\label{proof:sensing_opt}
\begin{proof}
	From \textit{Corollary \ref{corollary:S-opt-correlation}}, the sensing-optimal correlation matrix is deterministic, i.e., $\R_{\X}=\tilde{\Rr}_{\X}$. Hence,\footnote{Although the MMSE-minimization waveform has been studied, e.g., in \cite{MIMO-Radar-waveform}, there are major differences in that i) the transmitted signal $\X$ is strictly random, resulting in the stochastic MMSE-minimization problem \eqref{eqn:sensing_opt}, ii) the correlation at sensing Rx side, e.g., $\covr$, is not considered in \cite{MIMO-Radar-waveform} and iii) the correlation matrix ${\XX}^\herm{\XX}=T\I_{\Ns}\otimes{\R_{\X}}$ in \eqref{eqn:sensing_metric} has block-Toeplitz structure, which is neglected and yields an unachievable lower bound on the MMSE in \cite{MIMO-Radar-waveform}.}
	\begin{align*}
		\Es
		=\ \inf_{\tilde{\Rr}_{\X}}\ {{\tr}}\left[\left(\cov^{-1}+\frac{T}{\sigma_{\rm s}^2}\I_{\Ns}\otimes\tilde{\Rr}_{\X} \right)^{-1}\right],
	\end{align*}
	with the power constraint $\tr(\tilde{\Rr}_{\X}) = N\Po$. Denote $\F=\Ut^\herm\X/\sigma_{\rm s}$, it immediately follows $\X=\sigma_{\rm s}\Ut\F$. By further denoting $\bar\Fr:=\F\F^\herm$ and from the decomposition (\ref{eqn:eigen_decomosition_G}),
	\begin{equation*}
		\begin{aligned}
			&\ {\tr}\Big[(\cov^{-1}+\sigma_{\rm s}^{-2}\I_{\Ns}\otimes\X\X^\herm)^{-1}\Big]\\
			=&\ {\tr}\left[\left(\sigma_{\mathsf g}^{-2} \EIGr^{-1}\otimes\EIGt^{-1}+\I_{\Ns}\otimes\bar\Fr\right)^{-1}\right]\\
			=&\ \sum_{i=1}^{\Ns}\tr\Big[(\sigma_{\mathsf g}^{-2}\eigr{i}^{-1}\EIGt^{-1}+\bar\Fr)^{-1}\Big].
		\end{aligned}
	\end{equation*}

Constructing the Lagrangian
\begin{equation*}
	\mathcal{L}:= \sum_{i=1}^{\Ns}\tr\Big[(\sigma_{\mathsf g}^{-2}\eigr{i}^{-1}\EIGt^{-1}+\bar\Fr)^{-1}\Big]+\ell\Bigg[\tr(\bar\Fr)- \frac{TNP_0}{\sigma_{\rm s}^2}\Bigg],
\end{equation*}
where $\ell$ is the Lagrange multiplier for power constraint. Taking the derivative of $\mathcal{L}$ with respect to $\bar\Fr$ and setting it to zero yield
\begin{equation}
	\label{eqn:opt_condition_F}
	\sum_{i=1}^{\Ns}(\sigma_{\mathsf g}^{-2}\eigr{i}^{-1}\EIGt^{-1}+\bar\Fr)^{-2}=\ell\I_{N}.
\end{equation}
The left-hand side of \eqref{eqn:opt_condition_F} must be diagonal since $\ell\I_{N}$ is diagonal.
Further note that $\EIGt$ is diagonal, hence $\bar\Fr$ must be diagonal. Denote $\bar\Fr={\rm diag}(\bar{f}_1,\cdots,\bar{f}_N)$ with $\bar{f}_i\geq0$ and $\sum_{i=1}^N \bar{f}_i=TNP_0/\sigma_{\rm s}^2$. From the condition \eqref{eqn:opt_condition_F}, $\varphi_i(\bar{f}_i,\ell) = 0$ must be satisfied for $1\leq i\leq N$, where $\varphi_i(\cdot)$ is defined in \eqref{eqn:waterfilling_equation}.

Note that $\bar{f}_i\geq 0$ must be satisfied as it represents power allocation. In this case, the Karush-Kuhn-Tucker conditions \cite[Chapter 5]{boyd2004convex} are applied to verify the solution
\begin{equation}
	\bar{f}_i^\star =\varrho_i^+(\ell)\\
\end{equation}
is indeed the optimal solution, where the parameter $\ell$ is to guarantee $\sum_{i=1}^{N} \bar{f}_i^\star = TNP_0/\sigma_{\rm s}^2$.
	
	Denote $\bar\Fr^{\star}=\diag(\f_1^\star,\cdots,\f_N^\star)$ as the optimal solution. %Note that $\FF=\F^\herm\F$ is invariant to post-multiplication of $\F$ by any proper orthonormal matrix. Therefore, 
	The optimal $\F^{\star}$ has a more general form $\F^{\star} = (\bar\Fr^{\star})^{1/2}\stiefel$, where $\stiefel\in\mathbb{C}^{N\times T}$ is any matrix with orthonormal rows, i.e., 
	$\stiefel\stiefel^\herm=\I_N$. The corresponding sensing-optimal waveform is thus $\X_{\rm s}=\sigma_{\rm s}\Ut\F^{\star}=\Ut(\sigma_{\rm s}^2\bar\Fr^{\star})^{1/2}\stiefel.$ By setting $\powers=\sigma_{\rm s}^2\bar\Fr^{\star}$, \eqref{eqn:sensing_opt_waveform} can be derived. The sensing-optimal MMSE is therefore
	\begin{equation*}
		\begin{aligned}
			\Es &= \sum_{j=1}^{\Ns}\tr\Big[(\sigma_{\mathsf g}^{-2}\eigr{j}^{-1}\EIGt^{-1}+\bar\Fr)^{-1}\Big] \\
			&= \sum_{j=1}^{\Ns}\sum_{i=1}^{N}(\sigma_{\mathsf g}^{-2}\eigr{j}^{-1}\eigt{i}^{-1}+\bar{f}^\star_i)^{-1}.
		\end{aligned}
	\end{equation*}
	This completes the proof.
\end{proof}

\section{Proof of Corollary \ref{corollary:sensing_opt_reduced}} \label{proof:sensing_opt_reduced}
\begin{proof}
	When $\covr=\I_{\Ns}$, $\Ur=\I_{\Ns}$ and $\eigr{j}=1$, $\forall j\in\{1,\cdots,\Ns\}$. Hence, \eqref{eqn:waterfilling_equation} reduces to
	\begin{equation*}
		\begin{aligned}
			\varphi_i(f,\ell) &= \sum_{j=1}^{\Ns}(\sigma_{\mathsf g}^{-2}\eigt{i}^{-1}+f)^{-2}-\ell \\
			&= \Ns(\sigma_{\mathsf g}^{-2}\eigt{i}^{-1}+f)^{-2}-\ell.
		\end{aligned}
	\end{equation*}
	Solving $\varphi_i\Big(\varrho_i(\ell),\ell\Big)=0$ yields
	$\varrho_i(\ell) = \sqrt{\frac{\Ns}{\ell}}-\frac{1}{\sigma_{\mathsf g}^2\eigt{i}}$. 
	Therefore, $[\powers]_{i,i}=\sigma_{\rm s}^2\varrho_i^+(\ell)=\Big(\eta_{\rm s}-\frac{\sigma_{\rm s}^2}{\sigma_{\mathsf g}^2 \eigt{i}}\Big)^{\!+}$, where $\eta_{\rm s}:=\sigma_{\rm s}^2\sqrt{\frac{\Ns}{\ell}}$ for simplicity. Likewise,
	\begin{equation*}
			\Es =\!\!\!\!\! \sum_{\substack{1 \le i \le N \\ 1 \le j \le \Ns}}\frac{\sigma_{\mathsf g}^2\eigt{i}}{1+\sigma_{\mathsf g}^2\eigt{i}\varrho^+_i(\ell)}= \Ns\!\!\!\!\sum_{1 \le i \le N}\frac{1}{\sigma_{\mathsf g}^{-2}\eigt{i}^{-1}+\varrho^+_i(\ell)},
	\end{equation*}
	which further simplifies to \eqref{eqn:sensing_opt_MMSE_reduced}. 
	
	From \eqref{eqn:estimator}, the MMSE estimator is
	\begin{equation*}
		\begin{aligned}
			\hat{\g}(\s,\X) &= [\sigma_{\rm s}^2(\sigma_{\mathsf g}^2\I_{\Ns}\otimes\covt)^{-1} \!+\! (\I_{\Ns}\otimes\X\X^\herm)]^{-1}(\I_{\Ns}\otimes\X^\herm)^\herm\s \\
			&= \I_{\Ns}\otimes \Big[\Big(\frac{\sigma_{\rm s}^2}{\sigma_{\mathsf g}^2}\covt^{-1}+\X\X^\herm\Big)^{-1}\X\Big] \s,
		\end{aligned}
	\end{equation*}
	where the equality holds from the property of the Kronecker product. Plugging the closed-form sensing-optimal waveform \eqref{eqn:sensing_opt_waveform} into the above equation yields
	\begin{equation*}
		\begin{aligned}
			\hat{\g}(\s,\X_{\rm s}) &= \I_{\Ns}\otimes \Big[\Big(\frac{\sigma_{\rm s}^2}{\sigma_{\mathsf g}^2}\covt^{-1}+\Ut\powers\Ut^\herm\Big)^{-1}\Ut\powers^{1/2}\stiefel\Big]\s\\
			&= \I_{\Ns}\otimes \Big[\Ut\Big(\frac{\sigma_{\rm s}^2}{\sigma_{\mathsf g}^2}\EIGt^{-1}+\powers\Big)^{-1}\Ut^\herm\Ut\powers^{1/2}\stiefel\Big]\s\\
			&=  \I_{\Ns}\otimes \Big[\Ut\Big(\frac{\sigma_{\rm s}^2}{\sigma_{\mathsf g}^2}\EIGt^{-1}+\powers\Big)^{-1}\powers^{1/2}\stiefel\Big]\s.\\
		\end{aligned}
	\end{equation*}
	
	Note that both $\EIGt^{-1}$ and $\powers$ are diagonal matrices. According to the property of diagonal matrices, $\Big(\frac{\sigma_{\rm s}^2}{\sigma_{\mathsf g}^2}\EIGt^{-1}+\powers\Big)^{-1}\powers^{1/2}$ is also diagonal, with its $i$-th entry being
	\begin{equation*}
		\left[\frac{\sigma_{\rm s}^2}{\sigma_{\mathsf g}^2\eigt{i}}+\left(\eta_{\rm s}-\frac{\sigma_{\rm s}^2}{\sigma_{\mathsf g}^2\eigt{i}}\right)^+\right]^{-1}\cdot \sqrt{\left(\eta_{\rm s}-\frac{\sigma_{\rm s}^2}{\sigma_{\mathsf g}^2\eigt{i}}\right)^+},
	\end{equation*}
	which further simplifies to $\tilde{p}_i$. 
\end{proof}

%% file: bio.tex
\begin{IEEEbiographynophoto}{Zi-Jie Wang}
	(Graduate Student Member, IEEE) received the Bachelor's degree in electronic and computer engineering from Shanghai Jiao Tong University, Shanghai, China, in 2020, and the Master's degree in electronic engineering (signal processing) from Nanyang Technological University, Singapore, in 2022. He is currently pursuing the joint Ph.D. degree in information and communication engineering from Shanghai Jiao Tong University and the National University of Singapore, Singapore. His research interests involve signal processing and information theory, with their applications in wireless sensing, communications, and integrated sensing and communications systems.
\end{IEEEbiographynophoto}

\begin{IEEEbiographynophoto}{Xudong Wang}
	(Fellow, IEEE) received the PhD degree in electrical and computer engineering from
	the Georgia Institute of Technology in 2003. He is currently a professor with the thrust of Internet of Things and the dean of the College of Future Technology in Hong Kong University of Science and Technology (Guangzhou). He is also an affiliate professor with the Department of Electrical and Computer Engineering, University of Washington.
	He was the John Wu and Jane Sun chair professor with UM-SJTU Joint Institute, Shanghai Jiao Tong
	University. He was a senior research engineer, a senior network architect, and an
	R\&D manager with several companies. His research interests include wireless
	communication networks, distributed machine learning, edge computing, and
	joint communications and sensing. He was the editor of \emph{IEEE Transactions
		on Mobile Computing}, \emph{IEEE Transactions on Vehicular Technology}, \emph{Elsevier
		Ad Hoc Networks}, and \emph{China Communications}. He was the guest editor of
	several international journals. He was the general chair of 2017 IEEE 5G
	Summit in Shanghai and the TPC co-chair of the 32nd International Conference
	on Information Networking. He was the demo co-chair of ACM International
	Symposium on Mobile Ad Hoc Networking and Computing (ACM MOBIHOC
	2006), technical program co-chair of Wireless Internet Conference (WICON)
	2007, and general co-chair of WICON 2008. He was a voting member of IEEE
	802.11 and 802.15 Standard Committees.
\end{IEEEbiographynophoto}

\begin{IEEEbiographynophoto}{Giuseppe Caire} (S '92 -- M '94 -- SM '03 -- F '05)  was born in Torino in 1965. He received a
B.Sc. in Electrical Engineering  from Politecnico di Torino in 1990,  an M.Sc. in Electrical Engineering from Princeton University in 1992, and a Ph.D. from Politecnico di Torino in 1994.  He has been a post-doctoral research fellow with the European Space Agency (ESTEC, Noordwijk, The Netherlands) in 1994-1995, Assistant Professor in Telecommunications at the Politecnico di Torino, Associate Professor at the University of Parma, Italy,  Professor with the Department of Mobile Communications at the Eurecom Institute,  Sophia-Antipolis, France, a Professor of Electrical Engineering with the Viterbi School of Engineering, University of Southern California, Los Angeles, and he is currently an Alexander von Humboldt Professor with the Faculty of Electrical Engineering and Computer Science at the Technical University of Berlin, Germany. He is a member of the German National Academy of Sciences (Leopoldina) since 2024, and a member of the American National Academy of Engineering (NAE) since 2026. He received the Jack Neubauer Best System Paper Award from the IEEE Vehicular Technology Society in 2003,  the
IEEE Communications Society and Information Theory Society Joint Paper Award in 2004, in 2011, and in 2025, 
the Okawa Research Award in 2006,   the Alexander von Humboldt Professorship in 2014, the Vodafone Innovation Prize in 2015, an ERC Advanced Grant in 2018,  the Leonard G. Abraham Prize for best IEEE JSAC paper in 2019, the IEEE Communications Society Edwin Howard Armstrong Achievement Award in 2020, the 2021 Leibniz Prize  of the German National Science Foundation (DFG), and the  CTTC Technical Achievement Award of the IEEE Communications Society in 2023.  Giuseppe Caire is a Fellow of IEEE since 2005.  He has served in the Board of Governors of the IEEE Information Theory Society from 2004 to 2007, and as officer from 2008 to 2013. He was President of the IEEE Information Theory Society in 2011. 
His main research interests are in the field of communications theory, information theory, channel and source coding
with particular focus on wireless communications.   
\end{IEEEbiographynophoto}